\documentclass[12pt]{article}
\usepackage{graphicx}
\usepackage{subfigure}
\usepackage{placeins}
\usepackage{amsmath}
\usepackage{array}
\usepackage{multirow}
\usepackage{amssymb}
\usepackage{verbatim}
\usepackage{authblk}

\newcommand{\tr}{{\rm tr}}
\newcommand{\tauint}{\tau_{\rm int}}
\newcommand{\tauexp}{\tau_{\rm exp}}
\newcommand{\taudiff}{\tau_{\rm diff}}
\newcommand{\tautunn}{\tau_{\rm tunn}}
\newcommand{\taumeas}{\tau_{\rm meas}}
\newcommand{\schro}{Schr\"odinger}
\newcommand{\refeq}[1]{Eq.~(\ref{#1})}

\numberwithin{equation}{section}

\title{Diffusion of topological charge in lattice QCD simulations}

\author{Greg McGlynn\thanks{gem2128@columbia.edu} }
\author{Robert D. Mawhinney\thanks{rdm@phys.columbia.edu}}
\affil{Physics Department, Columbia University, New York, NY 10027, USA}

\date{\vspace{-5ex}}

\begin{document}
\maketitle

\begin{abstract}

We study the autocorrelations of observables constructed from the topological
charge density, such as the topological charge on a time slice or in a
subvolume, using a series of hybrid Monte Carlo simulations of pure SU(3) gauge
theory with both periodic and open boundary conditions. We show that the
autocorrelation functions of these observables obey a simple diffusion equation
and we measure the diffusion coefficient, finding that it scales like the
square of the lattice spacing. We use this result and measurements of the rate
of tunneling between topological charge sectors to calculate the scaling
behavior of the autocorrelation times of these observables on periodic and open
lattices.  There is a characteristic lattice spacing at which open boundary
conditions become worthwhile for reducing autocorrelations and we show how this
lattice spacing is related to the diffusion coefficient, the tunneling rate,
and the lattice Euclidean time extent.

\end{abstract}

\section{Introduction}

It is well known that in hybrid Monte Carlo (HMC) simulations of lattice QCD
the autocorrelation time of the topological charge increases very rapidly as
the lattice spacing is reduced \cite{Alpha, ThetaDep, LuscherProc, LuscherSF}.
This is understood to be a consequence of the fact that in a periodic volume
the topological charge of a continuum gauge field cannot change by any
continuous deformation, while the topological charge of a lattice gauge field
can only change by passing through non-continuum-like configurations with large
values of the action. As the coupling is made weaker such configurations are
more and more strongly suppressed, so that eventually ``tunneling'' between the
topological sectors of field space becomes very rare.

The resulting increase in the autocorrelation time of the topological charge is
dangerous, because when autocorrelation times become comparable to or longer
than the total length of a simulation there is no guarantee that the
statistical errors on measured quantities can be reliably estimated. The whole
calculation then becomes suspect. Modern simulations of QCD are being performed
at lattice spacings fine enough that this problem is a real and pressing one.

In \cite{LS_OBC}, it was proposed that switching from periodic to open boundary
conditions for the Euclidean time direction should slow the increase of the
autocorrelation time of the topological charge. The reason is that when open
boundary conditions are used topological charge can flow into or out of the
lattice through the boundaries and thus the topological charge can change
continuously without the need for rare tunneling events. Ref.~\cite{LS_OBC}
provided evidence for this hypothesis by studying the dependence of
autocorrelation times on the lattice spacing $a$ in simulations of pure SU(3)
gauge theory with open boundary conditions. Autocorrelation times were observed
to scale like $1/a^2$ at fine enough lattice spacings, which is a slower
increase than expected with periodic boundary conditions.  However, that work
was not able to make a direct comparison between periodic and open boundary
conditions because only open boundaries were simulated.

The present authors attempted such a comparison in \cite{MMProc} and did not
find any dramatic improvement from switching to open boundary conditions, but
the small statistics of that study made it impossible to draw precise
conclusions. That motivated this work, in which we collect very high statistics
and carry out a systematic comparison of periodic and open boundary conditions
across a wide range of lattice spacings. There has not yet been such a
systematic study, although some smaller-scale comparisons have been made
\cite{LuscherProc, OpenSusceptibility, NfDependence} and a similar study was
recently done in the context of the \schro{} functional \cite{LuscherSF}.

A systematic direct comparison between periodic and open boundaries is needed
because open boundary conditions have some drawbacks: they distort the physics
in the region of the lattice immediately adjacent to the boundaries and they
also break time-translational symmetry. These effects can be avoided by working
far from the lattice boundaries, but this requires sacrificing some of the
lattice volume to boundary effects. It is therefore important to find out
under what circumstances open boundary conditions can produce a worthwhile
reduction in autocorrelation times compared to traditional periodic boundary
conditions. 


In answering this question, we do more than provide raw numerical data on
autocorrelation times. We focus on observables constructed from the topological
charge density (which we will call ``topological observables'') and we show
that their autocorrelation functions can be reproduced by a simple mathematical
model that postulates only two processes: a tunneling process and a diffusion
process. This model fits our data surprisingly well and provides insight into
how the topological charge density evolves during the HMC algorithm. For
example, the model will tell us how quickly topological charge moves into the
lattice after being created at an open boundary.  

The free parameters of the model are the tunneling rate and the diffusion
coefficient. We measure the scaling of these parameters with $a$ and then use
this knowledge to compute the scaling behavior of topological autocorrelation
times. In the end, the model provides a criterion for deciding when open
boundary conditions are useful for reducing autocorrelation times.

This paper is organized as follows. In Section \ref{sec:SimDetails} we describe
the numerical simulations that form the basis of this work. In Section
\ref{sec:Results} we discuss which observables should be used in comparisons
between periodic and open boundary conditions and give the measured integrated
autocorrelation times of these observables as a function of the lattice
spacing. In Section \ref{sec:Diffusion} we develop our mathematical model for
topological autocorrelation functions, compare it to the data, and derive its
predictions for the scaling behavior of autocorrelation times.

\section{Numerical simulations} \label{sec:SimDetails}

In this section we describe the parameters of our simulations and define the
observables that we will study in later sections.

\subsection{Ensembles}

We simulate pure SU(3) gauge theory using the DBW2 gauge action \cite{DBW2},
which is defined by 
\begin{equation} \label{eq:DBW2} S_g = -\frac{\beta}{3} [(1 - 8 c_1) P + c_1
R], \,\,\,\, c_1 = -1.4088 \end{equation}
where $P$ is the sum of all unoriented $1 \times 1$ plaquettes and $R$ is the
sum of all unoriented $1 \times 2$ rectangles. For our purposes, the advantage
of the DBW2 action is that it lets us study the effects of nearly-frozen
topology at relatively coarse lattice spacings \cite{DBW2SlowdownRBC}. Already
at $a = 0.1$ fm the topological charge has autocorrelations of thousands of
molecular dynamics (MD) time units (MDU). To study such long autocorrelations
with, for example, the Wilson gauge action would require going to $a \sim 0.05$
fm. By allowing us to study the freezing of topology on relatively coarse
lattices, the DBW2 action lets us save computing resources by using relatively
small lattice volumes for a given physical volume. 

In the case of open boundary conditions there is some freedom to choose the
details of the action at the temporal boundaries. We make the following simple
choice: the action is given by \refeq{eq:DBW2} except that any plaquette or
rectangle which extends beyond one of the temporal boundaries is omitted from
the action. In our conventions the temporal boundaries are at Euclidean times
$t=0$ and $t=T-a$ (so the lattice comprises $N_t = T/a$ time slices).

Table \ref{tab:SimParameters} summarizes the parameters of our simulations,
which span a factor of two in lattice spacing. Our lattices all have physical
spatial extent $L = 1.6$ fm, with lattice volumes ranging from $8^3$ to $16^3$.
The Euclidean time extent $T$ of our lattices is always twice the spatial
extent. At the coarsest lattice spacings, topological tunneling is very
frequent, while at the finest lattice spacings topology is nearly frozen and
autocorrelation times are extremely long. We have collected enough statistics
to accurately measure these long autocorrelations even on the finest lattices.  

Each row of Table \ref{tab:SimParameters} represents two simulations: one with
periodic boundary conditions and one with open boundary conditions. The $\beta
= 0.9465$ row is an exception: for this lattice spacing we generated 4
independent ensembles for each boundary condition, for a total of 8 ensembles
at this lattice spacing (this was simply a convenient strategy given the
computer resources we used). All of our results at this lattice spacing are
averages over these sets of four independent ensembles.  

Following \cite{LS_OBC}, we scale the molecular dynamics trajectory
length\footnote{We use the conventions in \cite{MDConvention} to define MD
time. Other conventions exist which differ by a factor of $\sqrt{2}$. In
particular, a unit-length MD trajectory in our conventions is longer by a
factor of $\sqrt{2}$ than a unit-length trajectory in \cite{LS_OBC}.}
$\tau_{\rm traj}$ like $1/a$ and take measurements at an MD time interval
$\taumeas$ which we scale approximately like $1/a^2$. We perform the molecular
dynamics integration with a force gradient integrator \cite{ForceGradKennedy,
ForceGradHantao} and choose step sizes that lead to $>90\%$ acceptance rates
for all ensembles. For each pair of ensembles we find identical acceptance
rates for periodic and open boundary conditions.

\begin{table} \centering
\begin{tabular}{r|r|r|r|r|r|r|r}
$\beta$ & $a$ (fm) & Volume & $\tau_{\rm traj} $ & $N_{\rm steps}$ &  $\taumeas$ & MD time & Acc. \\
\hline
0.7796 & 0.2000(20) & $8^3 \times 16$  & 1.00 & 8 & 10 & 48410 & 
94%
\% \\
0.8319 & 0.1600(16) & $10^3 \times 20$ & 1.25 & 12 & 15 & 157875 & 
95%
\% \\
0.8895 & 0.1326(13) & $12^3 \times 24$ & 1.50 & 15 & 21 & 228438 & 
93%
\% \\
0.9465 & 0.1143(11) & $14^3 \times 28$ & 1.75 & 20 & 28 & 510944 & 
93%
\% \\
1.0038 & 0.1000(10) & $16^3 \times 32$ & 2.00 & 24 & 40 &  830560 & 
93%
\% \\
\hline
\end{tabular}
\caption{Simulation parameters. The lattice spacings in this table are computed
using Eq.~(4.11) of \cite{DBW2Spacing}, which gives $r_0/a$ as a function of
$\beta$ for the DBW2 action; we take $r_0 = 0.5$ fm and estimate a 1\%
statistical error based on the data in \cite{DBW2Spacing}. $\tau_{\rm traj}$ is
the HMC trajectory length in MDU, and each trajectory consists of $N_{\rm
steps}$ steps of the force gradient integrator.  $\taumeas$ is the MD time
separation between successive measurements of the observables described in
Section \ref{sec:Observables}. The listed MD time is the total length of the
simulation in MDU (for $\beta = 0.9465$ we ran four simulations of equal length
for each type of boundary condition and the listed MD time is the sum of the
lengths of the four simulations). The last column is the acceptance rate, which
we found to be independent of the boundary conditions.}
\label{tab:SimParameters}
\end{table}

\subsection{Observables} \label{sec:Observables}

The basic observables we study are the sums of the topological charge density
over single time slices, which we call $Q(t)$:
\begin{equation} \label{eq:Qslice} Q(t) \equiv a^4 \sum_{\vec x} \rho(\vec x,
t) \end{equation}
In the continuum the topological charge density $\rho$ is 
\begin{equation} \rho(\vec x, t) = \frac{1}{32 \pi^2}
\epsilon_{\mu\nu\rho\lambda} \tr (F_{\mu \nu}(\vec x, t) F_{\rho \lambda}(\vec
x, t)) \end{equation}
On the lattice we use the ``5Li'' discretization of this formula, defined in
\cite{5Li}. We always measure the topological charge density after smearing the
gauge field by running the Wilson flow to the reference flow time $t_0$
\cite{LuscherWflow}.

From the time slice observables $Q(t)$ we can also construct observables on 4D
subvolumes. As we will see, the charge summed over a large subvolume has a
longer autocorrelation time than the charge summed over a single time slice. We
define the topological charge $Q(t_1, t_2)$ summed over the Euclidean time
interval $[t_1, t_2)$ by
\begin{equation} \label{eq:Qsubvol} Q(t_1, t_2) \equiv \sum_{t_1 \leq t < t_2}
Q(t) \end{equation}
A particularly important special case is the ``global'' topological charge
summed over the entire lattice, $Q \equiv Q(0, T)$. 

In our discussion of boundary effects in Section \ref{sec:BoundaryEffects} we
will also consider one observable unrelated to topology: $E(t)$, the Yang-Mills
action density averaged over a single time slice, given by
\begin{equation} E(t) \equiv \frac{a^3}{L^3} \sum_{\vec x} \frac{1}{2} \tr
(F_{\mu \nu}(\vec x, t) F_{\mu \nu}(\vec x, t)) \end{equation}
In this formula we use the ``clover'' discretization of the field strength
tensor $F_{\mu\nu}$ \cite{LuscherWflow}. As with the topological charge
density, we measure the action density after running the Wilson flow to the
reference flow time $t_0$.

\section{Results} \label{sec:Results}

\subsection{Boundary effects} \label{sec:BoundaryEffects}

\begin{figure} \centering 
\includegraphics[width=3.8in]{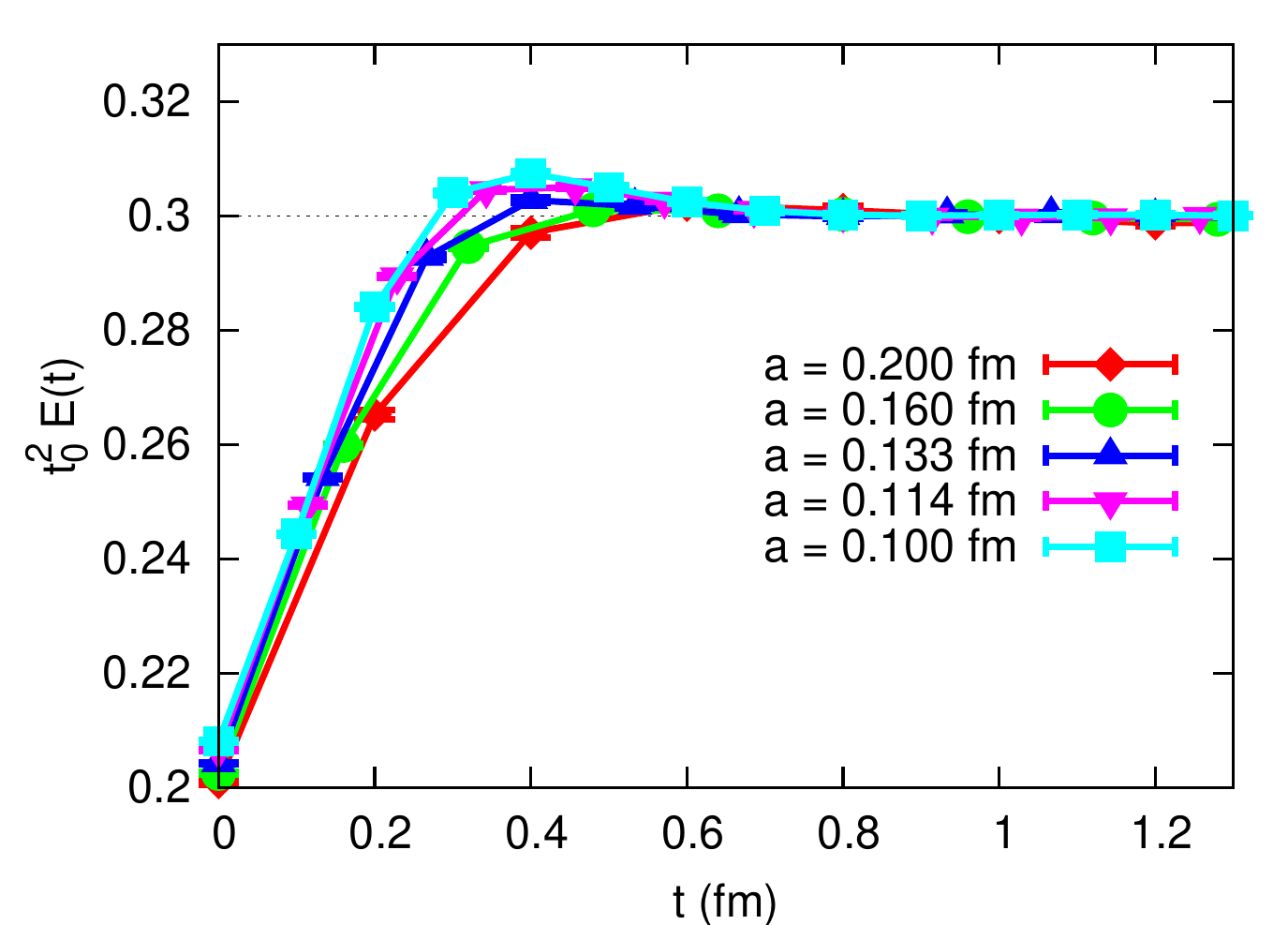}
\caption{Measurements of $t_0^2 \langle E(t) \rangle$ as a function of the
Euclidean time $t$ on open lattices near the $t = 0$ Euclidean time boundary.
Error bars are too small to see.}
\label{fig:ActionSliceOpenEdge}
\end{figure}

When open boundary conditions are used in a lattice QCD simulation, there is a
region near each open boundary in which the simulated physics is very different
from infinite-volume QCD. For example, Figure \ref{fig:ActionSliceOpenEdge}
shows the dimensionless quantity $t_0^2 \langle E(t) \rangle$ on open lattices
as a function of the Euclidean time $t$ near the open boundary at $t=0$. The
definition of $t_0$ is such that the true value of this observable is exactly
$0.3$ in an infinite volume, but it is evident that in the immediate vicinity
of $t=0$ the action density is quite different from its value in the central
region of the lattice. 

Similar boundary effects will be present in all observables. Ultimately, the
physics we are interested in is infinite-volume QCD, which means the physics in
the central region of Euclidean time, where measurements are independent of the
boundary conditions. We will call the central region of the lattice where the
physics is independent of the boundary conditions the ``bulk,'' in contrast to
the ``boundary regions'' near $t=0$ and $t=T-a$ where open and periodic
boundary conditions show significant differences. Figure
\ref{fig:ActionSliceOpenEdge} suggests that on our lattices a safe estimate for
the width of the boundary region is $T/4 = 0.8$ fm. We also examined boundary
effects in observables constructed from $Q(t)$, such as $\langle Q(t)^2
\rangle$, and found the boundary region to be no wider than that for $\langle
E(t) \rangle$. The width of the boundary region is presumably determined by a
combination of QCD correlation lengths (inverse glueball masses, in the pure
gauge theory) and the smearing radius $\sqrt{8 t_0} \sim 0.5$ fm of the Wilson
flow \cite{LuscherWflow}. The boundary region is likely narrower for observables
defined without smearing.

The fact that the physics is altered in the boundary regions means that a
useful comparison between periodic and open boundary conditions requires some
care. The boundary regions on an open lattice are simulating physics which is
not infinite-volume QCD and which has no analogue on a corresponding periodic
lattice. Therefore it is not sensible to include the boundary regions in any
comparison between an open ensemble and a periodic ensemble. For example, we
will not compare the autocorrelations of the global topological charge $Q$ on
periodic and open lattices because this observable contains large contributions
from the boundary regions. It turns out that autocorrelation times tend to be
much shorter in the boundary regions than in the bulk, so observables with
contributions from boundary regions will show artificially low autocorrelation
times on open lattices compared to periodic lattices. But when the goal is to
simulate infinite-volume QCD, this effect does not represent a speedup because
it comes from regions of the open lattice where the physics is very different
from infinite-volume QCD. The interesting question is whether autocorrelation
times in the \emph{bulk} are reduced by using open boundary conditions.
Therefore when we discuss autocorrelations we will only make comparisons
between open and periodic lattices using observables defined within the central
region $[T/4, 3T/4)$ of Euclidean time, which we found above to have
boundary-independent physics.

\subsection{Measured autocorrelations of topological observables} 
\label{sec:MeasAutocors}

In this section we give some measurements of autocorrelations of topological
observables on our ensembles. First we briefly clarify our conventions for
measures of autocorrelation. Suppose we measure some observable $X$ as a
function of MD time $\tau$. Then $\Gamma_X$, the autocorrelation function of
$X$, is defined as
\begin{equation} \Gamma_X(\tau) = \langle X(\tau_0 + \tau) X(\tau_0) \rangle -
\langle X \rangle^2 \end{equation}
The normalized autocorrelation function $\rho_X(\tau)$ and integrated
autocorrelation time $\tauint(X)$ are defined by 
\begin{equation} \rho_X(\tau) = \frac{\Gamma_X(\tau)}{\Gamma_X(0)}
\,\,\,\,\,\,\,\,\,\,\,\, \tauint(X) = \frac{\taumeas}{2} \sum_{n =
-\infty}^\infty \rho_X(n \taumeas) \end{equation}
where $\taumeas$ is the MD time interval at which we measure $X$. We always
report integrated autocorrelation times in molecular dynamics time units.
Ref.'s~\cite{Alpha, LS_OBC} contain useful formulas for calculating statistical
errors on the estimators of these quantities.

\begin{figure} \centering
\includegraphics[width=5.3in]{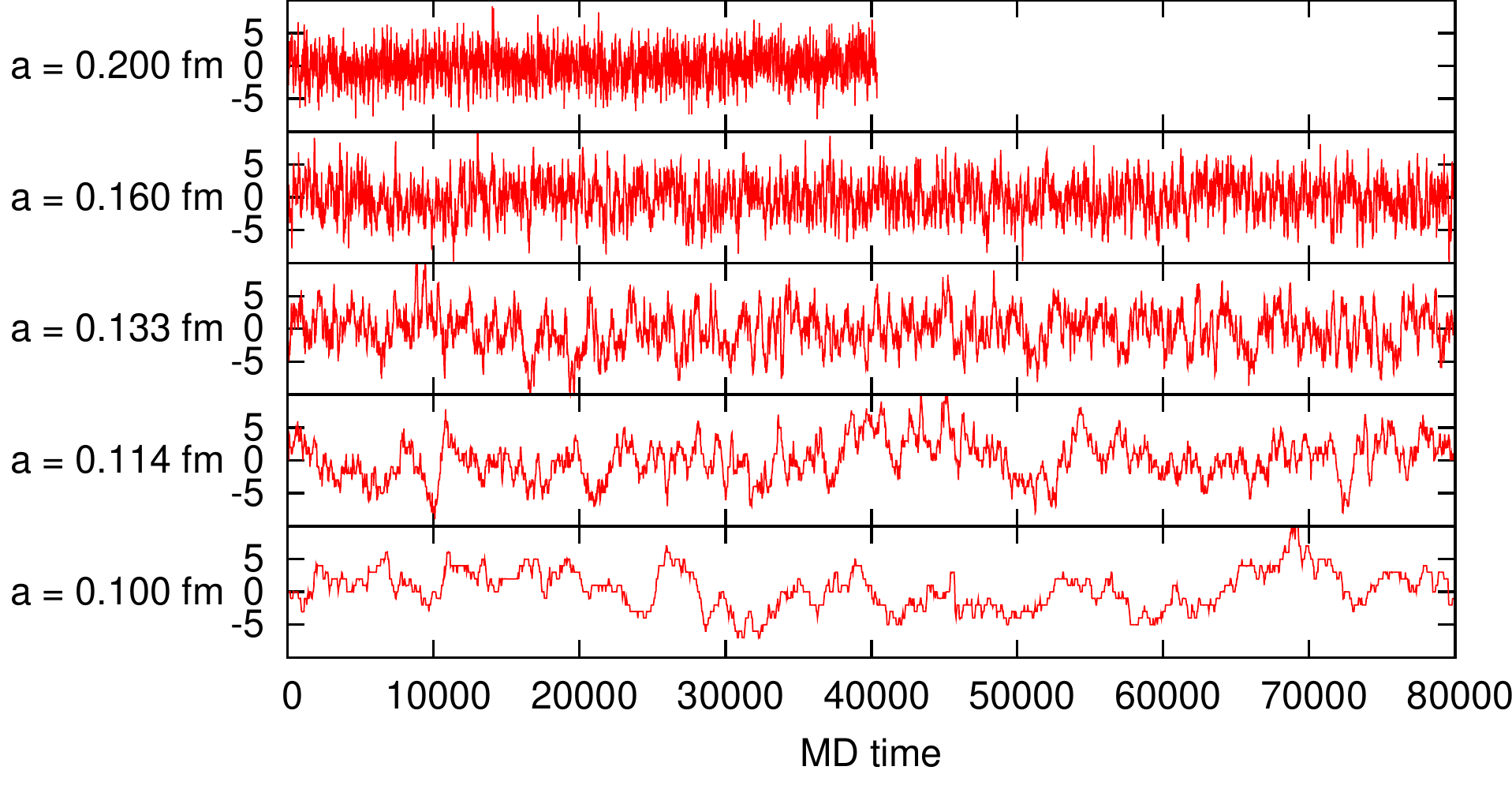}
\caption{Portions of the MD time histories of the global topological charge $Q$
from periodic ensembles.}
\label{fig:GlobalQHists}
\end{figure}

\begin{figure} \centering
\includegraphics[width=3.8in]{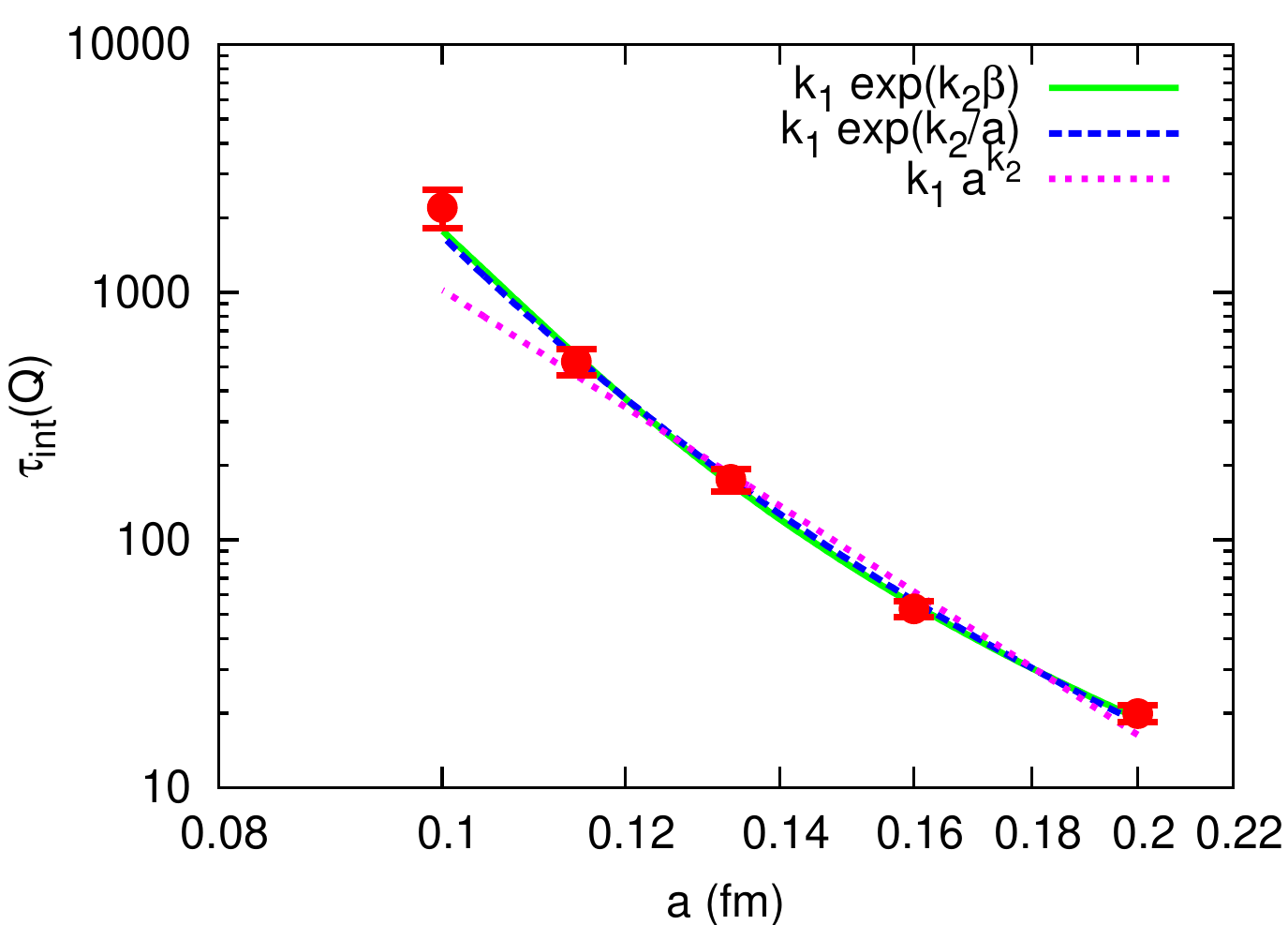}
\caption{Fits to the scaling behavior of $\tauint(Q)$ on periodic lattices. $a$
and $\beta$ are related using Eq. (4.11) of \cite{DBW2Spacing}.}
\label{fig:GlobalQIATs}
\end{figure}

As discussed in the introduction, the global topological charge $Q$ rapidly
develops longer and longer autocorrelations as the lattice spacing is
decreased. In Figure \ref{fig:GlobalQHists} we show portions of the MD time
histories of $Q$ on our periodic lattices at each simulated lattice spacing.
The dramatic slowdown of $Q$ as $a \to 0$ is obvious. Table
\ref{tab:SubvolumeIATs} gives $\tauint(Q)$, the integrated autocorrelation time
of the global topological charge, on each of our periodic lattices.
$\tauint(Q)$ increases by a factor of about 100 from our coarsest to our finest
lattice. Figure \ref{fig:GlobalQIATs} shows that we obtain a good fit to the
scaling behavior of $\tauint(Q)$ with the ansatz
\begin{equation} \label{eq:QGlobalExpBeta} \tauint(Q) = k_1 e^{k_2 \beta},
\,\,\,\,\,\,\, k_1 = 2.7(1.7) \times 10^{-6} \text{ MDU}, \,\,\,\,\,\,\, k_2 =
20.2(7) \end{equation} 
This fit form is motivated by the notion that there is some action barrier
$S_0$ to topological tunneling which should therefore be suppressed by a factor
$e^{-\beta S_0}$. We can also obtain a good fit using the form 
\begin{equation} \label{eq:QGlobalExpA} \tauint(Q) = k_1 \exp(k_2/a),
\,\,\,\,\,\,\, k_1 = 0.20(5) \text{ MDU}, \,\,\,\,\,\,\, k_2 = 0.90(3)\text{
fm} \end{equation}
A power law $k_1 a^{k_2}$ with $k_2 \approx -6$, can approximately fit the
data, but this fit is not as good, as Figure \ref{fig:GlobalQIATs} shows.

These results for the $a$-dependence of $\tauint(Q)$ are quite similar to those
of \cite{Alpha}, which simulated the pure gauge theory with the Wilson gauge
action, and found that both the form of \refeq{eq:QGlobalExpA} and the
power law form (with exponent around $-5$) described the data reasonably well.
While we use the DBW2 gauge action and so $\tauint(Q)$ becomes large at a
coarser lattice spacing, the same fit forms apparently work reasonably well for
both actions.

\begin{table} \centering
\begin{tabular}{c||c||c|c||c|c||}
       & $Q$      & \multicolumn{2}{c||}{$Q(T/2)$} & \multicolumn{2}{c||}{$Q(T/4,3T/4)$} \\
\cline{2-6}
$a$ (fm) & Periodic & Periodic & Open                & Periodic      & Open \\
\hline
0.2000 & 20(2) 
       & 13.0(4) 
       & 13(1) 
       & 18(1) 
       & 18(1) \\
0.1600 & 53(4) 
       & 28(1) 
       & 28(2) 
       & 44(2) 
       & 41(3) \\
0.1326 & 175(18) 
       & 66(4) 
       & 62(5) 
       & 129(11) 
       & 105(9) \\
0.1143 & 525(63) 
       & 151(11) 
       & 136(11) 
       & 353(35) 
       & 270(22) \\
0.1000 & 2197(389) 
       & 465(66) 
       & 217(19) 
       & 1307(214) 
       & 464(43) \\
\hline
\end{tabular}
\caption{Measured integrated autocorrelation times of some topological
observables. Open boundary conditions lead to significantly shorter integrated
autocorrelation times when the lattice spacing is fine enough. (However, even
these shorter times are still quite long.)}
\label{tab:SubvolumeIATs}
\end{table}

\begin{figure} \centering
\includegraphics[width=3.8in]{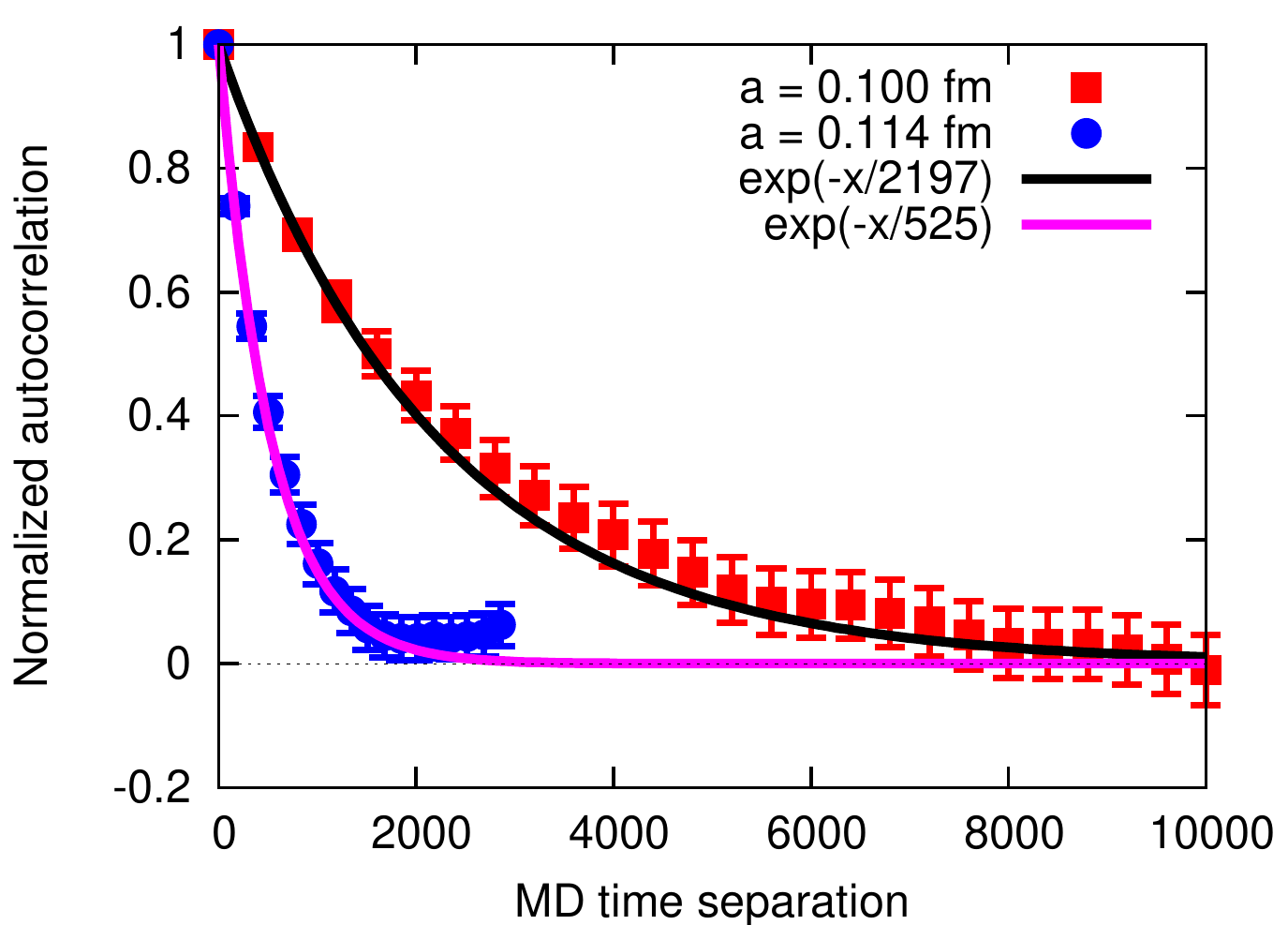}
\caption{Measured normalized autocorrelation functions of the global charge $Q$
on our two finest periodic lattices. Also shown are curves of the form
$\exp(-\tau/\tauint)$ where $\tauint$ is the measured integrated
autocorrelation time. This single-exponential form matches the measured
autocorrelation functions well.}
\label{fig:QglobalPureExp}
\end{figure}

On all of our periodic lattices we find that the autocorrelation function of
$Q$ has the form of a single exponential to within our statistical precision.
Figure \ref{fig:QglobalPureExp} shows this for our two finest lattice spacings.

As discussed in Section \ref{sec:BoundaryEffects}, the global topological
charge $Q$ is not the best observable to use for comparisons between periodic
and open lattices. We should instead look at observables defined on subvolumes
that lie entirely within the bulk. For the moment we focus on two such
observables: $Q(T/2)$, the topological charge summed over the central time
slice, and $Q(T/4, 3T/4)$, the topological charge summed over the central half
of the lattice volume. 

Table \ref{tab:SubvolumeIATs} gives the integrated autocorrelation times of
these observables on both periodic and open lattices at each lattice spacing.
Like $\tauint(Q)$, these integrated autocorrelation times rise very rapidly as
the lattice spacing is reduced. However, the $a$-dependence of these
autocorrelation times is not captured by a simple function like
\refeq{eq:QGlobalExpBeta} or (\ref{eq:QGlobalExpA}). We determine the scaling
behavior of these autocorrelation times in Section \ref{sec:Diffusion} below.
Note that the half-volume charge $Q(T/4, 3T/4)$ always has a significantly
longer autocorrelation time than the time-slice charge $Q(T/2)$. 

The results of Table \ref{tab:SubvolumeIATs} show that open boundary conditions
do indeed lead to reduced (but still quite long) autocorrelation times at fine
enough lattice spacings. At the very finest lattice spacing, $a = 0.1$ fm, open
boundary conditions produce a more than a factor of 2 reduction in the
integrated autocorrelation times. At $a = 0.114$ fm and 0.133 fm, the next two
finest lattice spacings, open boundary conditions show slightly shorter
autocorrelations than periodic boundary conditions, with the improvement
clearer for the half-volume charge $Q(T/4, 3T/4)$. At the two coarsest lattice
spacings the integrated autocorrelation times are independent of the boundary
conditions to within the limits of our measurements. 

\begin{figure} \centering
\includegraphics[width=5.3in]{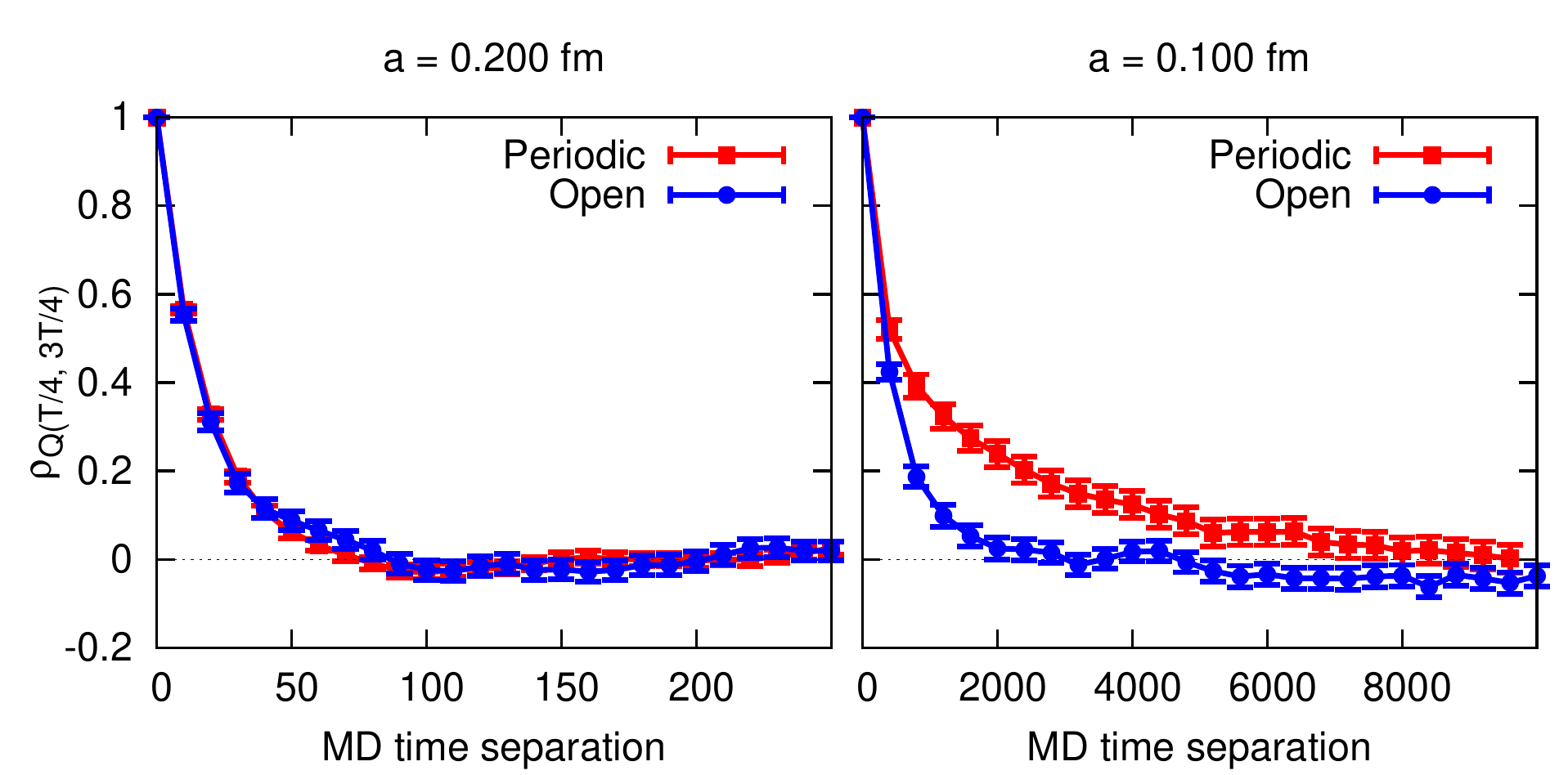}
\caption{Normalized autocorrelation function of $Q(T/4,3T/4)$, the topological
charge summed over the central half of the lattice time extent. At $a = 0.100$
fm, the autocorrelation function has a very long tail on the periodic lattice
which is absent on the open lattice.}
\label{fig:QhalfACFs}
\end{figure}

The reason for the difference between periodic and open boundary conditions at
fine lattice spacings is exactly that envisaged in \cite{LS_OBC}. At fine
lattice spacings on periodic lattices, the autocorrelation functions of
topological observables like the time-slice charge and the half-volume charge
develop long tails proportional to the autocorrelation function of the global
charge. As the autocorrelation time of the global charge becomes very long, so
do these tails. Autocorrelation functions on open lattices do not develop such
long tails, because the global charge does not slow down as drastically. On
open lattices the topological charge can flow in and out through the boundaries
and so the global charge can change without having to wait for rare tunneling
events. Figure \ref{fig:QhalfACFs} demonstrates this, comparing the
autocorrelation function of the half-volume charge $Q(T/4,3T/4)$ between open
and periodic boundary conditions at the coarsest and finest lattice spacings.

In the rest of this paper we will develop a model for topological
autocorrelations which will precisely reproduce the measured autocorrelation
functions of topological observables, such as those plotted in Figure
\ref{fig:QhalfACFs}. Among other things this model lets us predict the lattice
spacing at which integrated autocorrelation times on periodic lattices start to
become much longer than those on open lattices. Thus the model will tell us
when open boundary conditions start to become useful for reducing
autocorrelations.

\section{Diffusion of topological charge} \label{sec:Diffusion}

The mechanism by which the global topological charge changes during an HMC
evolution is moderately well-understood. As discussed in the introduction, the
global charge on a periodic lattice can only change via lattice artifacts:
``tears'' or ``dislocations'' in the gauge field where the field is not smooth
and continuum-like. These dislocations are likely to be small structures, with
size of order the lattice scale, in order to minimize their action. When the
global topological charge changes by means of one of these dislocations we
speak of the lattice gauge field tunneling between adjacent topological
sectors. The rate of tunneling can be quantified by, for example, the
integrated autocorrelation time of the global topological charge.

Less well-understood is how the topological charge moves around the lattice in
the absence of these tunneling events. In particular, when considering open
boundary conditions it would be very useful to know how fast this motion is,
because open boundary conditions are supposed to reduce autocorrelations by
allowing topological charge to be created or destroyed at the open boundaries
and then move into the bulk of the lattice. The effectiveness of open boundary
conditions will therefore be directly related to the speed at which topological
charge moves through the lattice in the absence of tunneling. (This question is
also interesting for simulations which deliberately run in a fixed topological
sector; then the rate at which charge moves around will determine how long it
takes the lattice to decorrelate within a given topological sector.) 

One of the strengths of the mathematical model we will now develop is that it
provides a clean and quantitative definition of the vague notion of ``how fast
topological charge moves around the lattice.'' This will enable us to develop a
theoretical understanding of the circumstances in which open boundary
conditions will reduce autocorrelations.

\subsection{The diffusion model}

In this section we give a mathematical model that reproduces the
autocorrelation function of $Q(t)$, the topological charge summed over a single
time slice. With this model we will be able to determine the scaling behavior
of the autocorrelation times of Section \ref{sec:MeasAutocors} and we will show
how to determine the lattice spacing at which the autocorrelation times
measured on open and periodic lattices start to differ.

Denote by $Q(t, \tau)$ the topological charge summed over the time slice with
Euclidean time coordinate $t$ on the configuration at MD time $\tau$. We will
focus on the correlation function\footnote{Henceforth ``$t_0$'' will always be
a Euclidean time and should not be confused with the Wilson flow reference
time.}
\begin{equation} \label{eq:DiffC} C(t, t_0, \tau) \equiv \langle Q(t, \tau_0 +
\tau) Q(t_0, \tau_0) \rangle \end{equation}
This correlation function tells us about the movement of topological charge
through the lattice during the HMC evolution. Roughly speaking, $C(t, t_0,
\tau)$ will be large when a lump of topological charge present on time slice
$t_0$ at some MD time $\tau_0$ is likely to move to time slice $t$ by MD time
$\tau_0 + \tau$. As a special case, $C(t_0, t_0, \tau)$ is the autocorrelation
function of $Q(t_0)$.

We can measure the correlation function $C$ straightforwardly with our high
statistics. We find empirically that it obeys a simple diffusion-decay
equation:\footnote{This generalization of the diffusion equation to a
position-dependent diffusion coefficient $D(t)$ is just one of many
possibilities.  This form is the first that we tried and we found it to work
well. Later we tried altering the diffusion term to $\partial^2/\partial
t^2(D(t)C)$. This form also works well but led to slightly larger values of the
$\chi^2$ defined in \refeq{eq:ChiSq}}
\begin{equation} \label{eq:DiffEq} \frac{\partial}{\partial \tau} C(t, t_0,
\tau) = \frac{\partial}{\partial t}\left(D(t) \frac{\partial}{\partial t} C(t,
t_0, \tau) \right) - \frac{1}{\tautunn} C(t, t_0, \tau) \end{equation}
Here the derivatives $\partial/\partial t$ with respect to Euclidean time
should be understood as finite differences and $D(t)$ is a function defined at
Euclidean times midway between the lattice time slices. We will call
\refeq{eq:DiffEq} the ``diffusion model.'' The free parameters of the model are
the function $D(t)$ and the quantity $\tautunn$.

$D(t)$ is a $t$-dependent diffusion coefficient with units of
$\text{fm}^2/\text{MDU}$. It quantifies how fast topological charge diffuses in
the Euclidean time direction and answers the question raised above of how fast
topological charge moves around the lattice in the absence of tunneling events.
By time-translation invariance, $D(t)$ is a constant function on periodic
lattices or in the bulk region of open lattices, but it can have nontrivial
$t$-dependence near open boundaries. In fact we will find in Section
\ref{sec:DiffusionFits} that $D(t)$ is somewhat enhanced in the immediate
vicinity of an open boundary. However, we will often treat $D(t)$ as a
constant, $D$, unless we are interested in this boundary effect.

$\tautunn$, which we call the ``tunneling timescale,'' has units of MD time and
quantifies the rate of tunneling between topological sectors. In fact, on a
periodic lattice it is exactly the integrated autocorrelation time of the
global topological charge. This can be seen as follows. Summing $C(t, t_0,
\tau)$ over $t$ and $t_0$ gives the autocorrelation function $\Gamma_Q$ of the
global topological charge:
\begin{equation} \Gamma_Q(\tau) \equiv \langle Q(\tau_0 + \tau) Q(\tau_0)
\rangle = \sum_{t=0}^{T-a} \sum_{t_0=0}^{T-a} C(t, t_0, \tau) \end{equation}
where here $Q(\tau)$ denotes the global topological charge at MD time $\tau$.
Then if we sum \refeq{eq:DiffEq} over $t$ and $t_0$, the diffusion term drops
out because it is a total derivative, leaving
\begin{equation} \label{eq:DiffModelGlobalQ} \frac{d}{d\tau} \Gamma_Q(\tau) =
-\frac{1}{\tautunn} \Gamma_Q(\tau) \end{equation}
This implies that the autocorrelation function of the global charge $Q$ is a
simple exponential, as found in Section \ref{sec:MeasAutocors}, and that the
area under the normalized autocorrelation function $\rho_Q(\tau)$ is
$\tautunn$, as claimed. Thus on periodic lattices we have $\tautunn =
\tauint(Q)$. 

In principle, $\tautunn$ could be a function of Euclidean time $t$ near an open
boundary. However, we are unable to resolve any such $t$-dependence in our
data, and so we always take $\tautunn$ to be a constant throughout the lattice.

The boundary conditions satisfied by the correlation function $C$ depend on the
boundary conditions for the lattice gauge field. On periodic lattices $C(t,
t_0, \tau)$ is periodic in $t$ and in $t_0$. On open lattices $C$ goes to zero
at the Euclidean time boundaries in the continuum limit:
\begin{equation} \label{eq:BCOpenC} C(0, t_0, \tau) = C(T-a, t_0, \tau) = C(t,
0, \tau) = C(t, T-a, \tau) = 0 \end{equation}
These boundary conditions let the correlations measured by $C$ ``leak out''
through the open boundaries, just as the topological charge itself can leak
out. They arise from the fact that open boundary conditions correspond to
setting the color-electric field $\vec{E}$ to zero at the boundaries
\cite{LS_OBC}. Therefore the topological charge density, which is proportional
to $\tr (\vec{E} \cdot \vec{B})$, also vanishes at the boundaries, as do
correlation functions of the charge density such as $C$.

The diffusion model provides a concrete way of thinking about how the
topological charge density changes during an HMC evolution. There are two
processes: a diffusion process that proceeds at a rate given by $D$ and a
tunneling process that proceeds at a rate given by $\tautunn$. As we will now
demonstrate, this simple model suffices to completely explain our measurements
of the autocorrelations of topological observables. The integrated
autocorrelation time $\tautunn$ is of course a well-known quantity but as far
as we know the diffusion coefficient $D$ has not been identified before.

\subsection{Diffusion model fits to simulation data} \label{sec:DiffusionFits}

In this section we discuss our method for estimating the free parameters of
\refeq{eq:DiffEq} from our data and demonstrate the close agreement between the
model and our simulation data.

\refeq{eq:DiffEq} predicts $C(t, t_0, \tau)$ for MD time separations $\tau > 0$
given the ``initial condition'' $C(t, t_0, \tau=0)$, which gives the
correlations between the $Q(t)$ at zero MD time separation. The free parameters
in the differential equation are the diffusion coefficient $D(t)$ and the
tunneling timescale $\tautunn$. $D(t)$ must be a constant function on periodic
lattices, but on open lattices we allow it to be a general function of $t$,
except that we impose time-reversal symmetry (\emph{i.e.}, symmetry under $t
\to T - a - t$). So on periodic lattices, the model has two free parameters,
$\tautunn$ and $D$, while on open lattices the model has $T/2+1$ free
parameters, $\tautunn$ and the values of $D(t)$ for $t < T/2$.

For a given choice of the parameters $D(t)$ and $\tautunn$, we define a measure
of the goodness of fit as follows. We measure the function $C(t, t_0, 0)$ from
our data, then numerically integrate \refeq{eq:DiffEq} to obtain the prediction
for the correlation function at $\tau > 0$, which we will denote by $C_{\rm
model}$.  We then measure the function $C(t, t_0, \tau)$ from our data,
obtaining an estimate $\bar C$ with statistical error $\Delta \bar C$. These
measurements are made for a discrete set of MD time separations $n \taumeas$,
$n = 1, 2, ..., N$.  Finally we define the goodness of fit
\begin{equation} \label{eq:ChiSq} \chi^2 \equiv \sum_{n=1}^{N} \sum_{t=0}^{T-a}
\sum_{t_0=0}^{T-a} \left( \frac{C_{\rm model}(t,t_0,n\taumeas) - \bar
C(t,t_0,n\taumeas)}{\Delta \bar C(t, t_0, n\taumeas)} \right)^2 \end{equation}

We find the best estimates of $D(t)$ and $\tautunn$ by varying them to minimize
this $\chi^2$. Finally we estimate statistical errors on $D(t)$ and $\tautunn$
by the jackknife method. We use a blocked jackknife with blocks much longer
than the longest measured autocorrelation time.

\begin{figure} \centering
\includegraphics[width=5.3in]{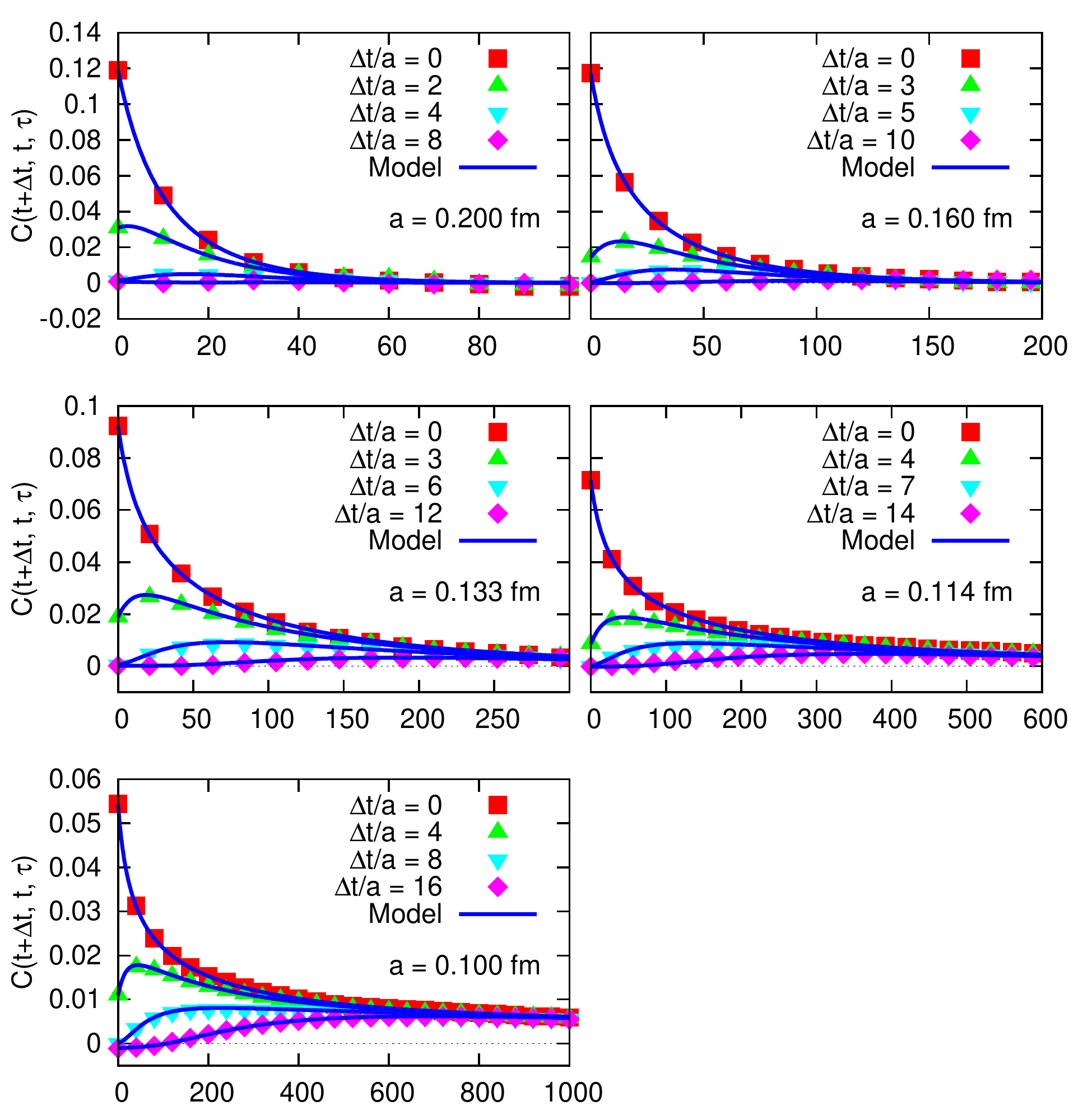}
\caption{Measurements and diffusion model predictions of $C(t+\Delta t, t,
\tau)$ vs $\tau$ for several values of $\Delta t$ on each periodic ensemble.
This is the correlation between the charge on time slice $t$ and the charge on
time slice $t+\Delta t$ an MD time $\tau$ later. By time translation invariance
this function is independent of $t$. Error bars on measurements are too small
to see. In all cases the model curve closely matches the measured data.}
\label{fig:CorrFitPeriodic}
\end{figure}

\begin{figure} \centering
\includegraphics[width=5.3in]{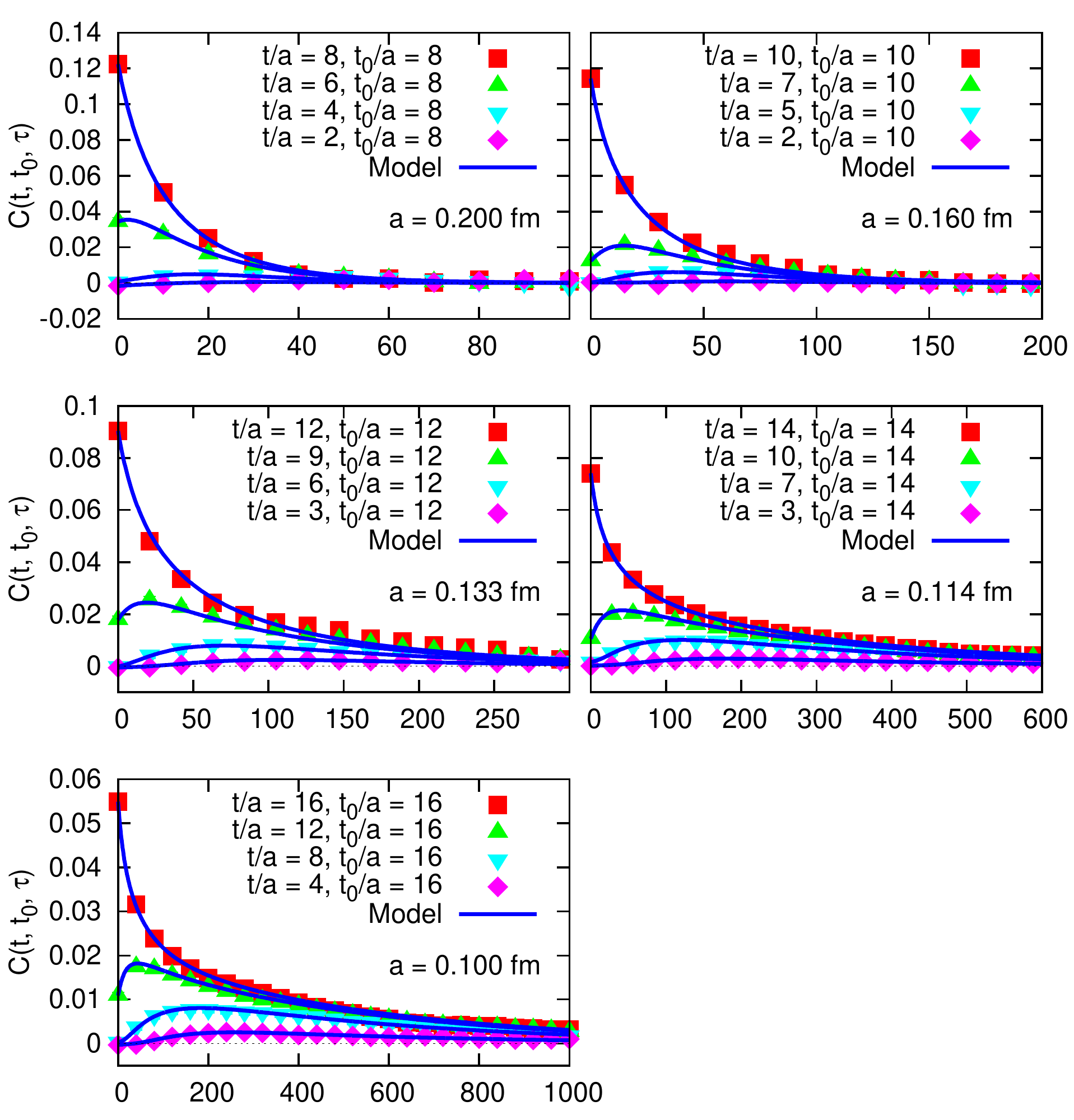}
\caption{Measurements and diffusion model predictions of $C(t, T/2, \tau)$ vs
$\tau$ for several values of $t$ on each open ensemble. This is the correlation
between the charge on the central time slice and the charge on time slice $t$
an MD time $\tau$ later. Error bars on measurements are too small to see. In
all cases the model curve closely matches the measured data.}
\label{fig:CorrFitOpen}
\end{figure}

The resulting fits are shown in Figures \ref{fig:CorrFitPeriodic} and
\ref{fig:CorrFitOpen} where for each ensemble we plot $C(t, t_0, \tau)$ for
several choices of $t$ and $t_0$ alongside the model fit. In every case our
simple model produces remarkably good agreement with the measured correlation
functions. We stress that $\tautunn$ and $D(t)$ are determined only once per
ensemble: the $\chi^2$ in \refeq{eq:ChiSq} sums over all values of $t$ and
$t_0$. After minimizing $\chi^2$, the resulting estimates for $\tautunn$ and
$D(t)$ are used to predict the $C(t, t_0, \tau)$ for any $t$ and any $t_0$.

\begin{figure} \centering
\includegraphics[width=3.8in]{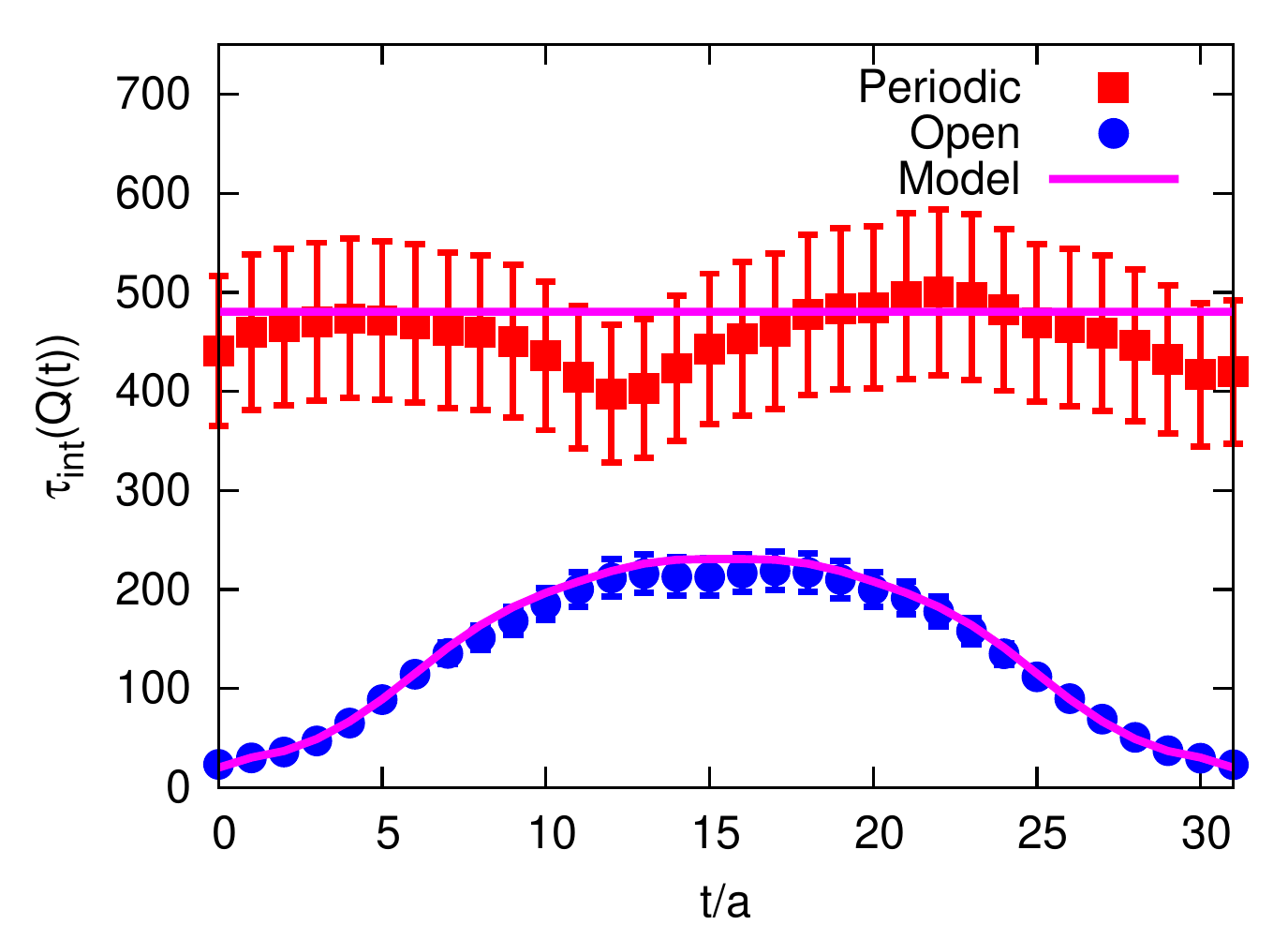}
\caption{Measurements and diffusion model prediction of $\tauint(Q(t))$ vs $t$
on the $a = 0.100$ fm periodic and open lattices.}
\label{fig:Qslice_IAT_vs_t}
\end{figure}

We can integrate the diffusion model predictions for autocorrelation functions
to get predictions for integrated autocorrelation times. Figure
\ref{fig:Qslice_IAT_vs_t} compares measurements of $\tauint(Q(t))$, the
integrated autocorrelation time of $Q(t)$, to model predictions on our finest
pair of ensembles. The predictions are computed using the estimates of $D(t)$
and $\tautunn$ from the above fitting procedure. There is close agreement, and
the diffusion model correctly reproduces the nontrivial $t$-dependence of
$\tauint(Q(t))$ in the presence of open boundary conditions.

The best estimates for $\tautunn$ and $D(\frac{T-a}{2})$ (the diffusion
coefficient at the center of the lattice) are summarized in Table
\ref{tab:ModelFitResults}. At a given lattice spacing, the measured values of
these quantities on the periodic and open lattices are consistent with each
other. This is expected: the boundary conditions should not affect the rate of
tunneling or diffusion in the bulk. Furthermore, the measured values of
$\tautunn$ in Table \ref{tab:ModelFitResults} are consistent with the measured
values of the integrated autocorrelation time $\tauint(Q)$ on periodic lattices
in Table \ref{tab:SubvolumeIATs}, as expected since \refeq{eq:DiffModelGlobalQ}
predicts $\tauint(Q) = \tautunn$. The estimates from open lattices tend to have
larger error bars: this is because the model has more free parameters on open
lattices since $D(t)$ is allowed to depend on $t$.

\begin{table} \centering
\begin{tabular}{c||c|c||c|c||}
       & \multicolumn{2}{c||}{$\tautunn$} & \multicolumn{2}{c||}{$D(\frac{T-a}{2})/a^2 \text{  (MDU}^{-1})$} \\
\cline{2-5}
$a$ (fm) & Periodic & Open                  & Periodic      & Open \\
\hline
0.2000 & 20(1) 
       & 20(2)
       & 0.090(12)
       & 0.099(30) \\
0.1600 & 56(3) 
       & 51(3)
       & 0.1018(73)
       & 0.113(18) \\
0.1326 & 185(20) 
       & 162(19)
       & 0.1085(97)
       & 0.088(15) \\
0.1143 & 561(59) 
       & 737(143)
       & 0.1080(56)
       & 0.120(14) \\
0.1000 & 2350(389) 
       & 1973(621)
       & 0.1155(29)
       & 0.116(12) \\
\hline
\end{tabular}
\caption{Best fit results for diffusion model parameters.}
\label{tab:ModelFitResults}
\end{table}

\begin{figure} \centering
\includegraphics[width=3.8in]{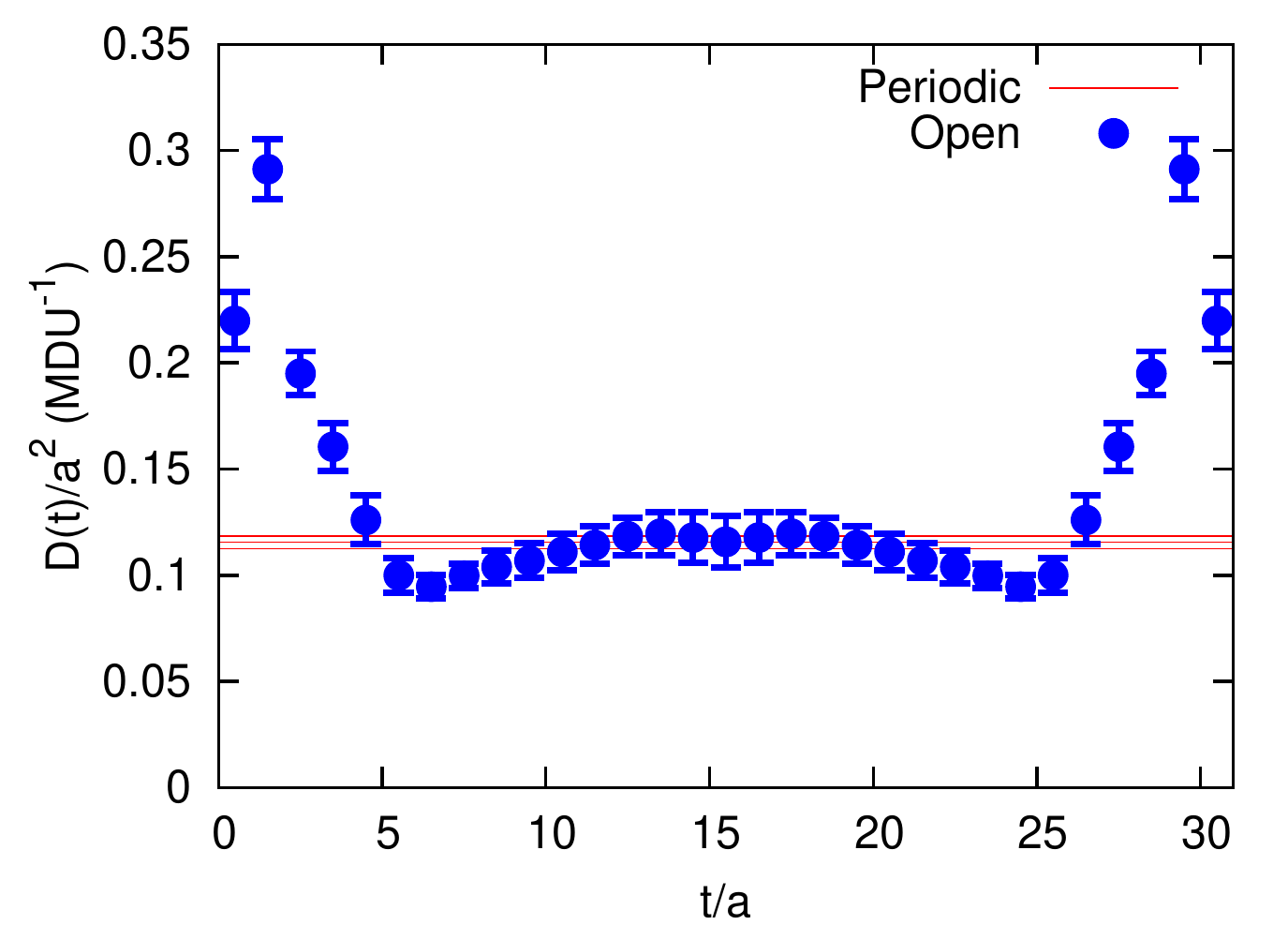}
\caption{Measured diffusion coefficient $D(t)$ on $a = 0.100$ fm lattices.}
\label{fig:D_vs_t}
\end{figure}

While $D(t)$ is identical in the bulk between periodic and open lattices, when
$t$ is close to an open boundary we observe that $D(t)$ is enhanced relative to
its bulk value. As an example, Figure \ref{fig:D_vs_t} shows the fit results
for the function $D(t)$ at our finest lattice spacing, comparing the open
result to the (time-translation invariant) periodic result.

$D(t)$ is a property of the HMC algorithm and not a physical observable.
However, we expect that the width of the boundary region in which $D(t)$ is
enhanced is controlled, as for physical observables, by a combination of the
Wilson flow smearing radius $\sqrt{8 t_0}$ and QCD correlation lengths.

\subsection{Scaling of the diffusion coefficient} \label{sec:DScaling}

\begin{figure} \centering
\includegraphics[width=3.8in]{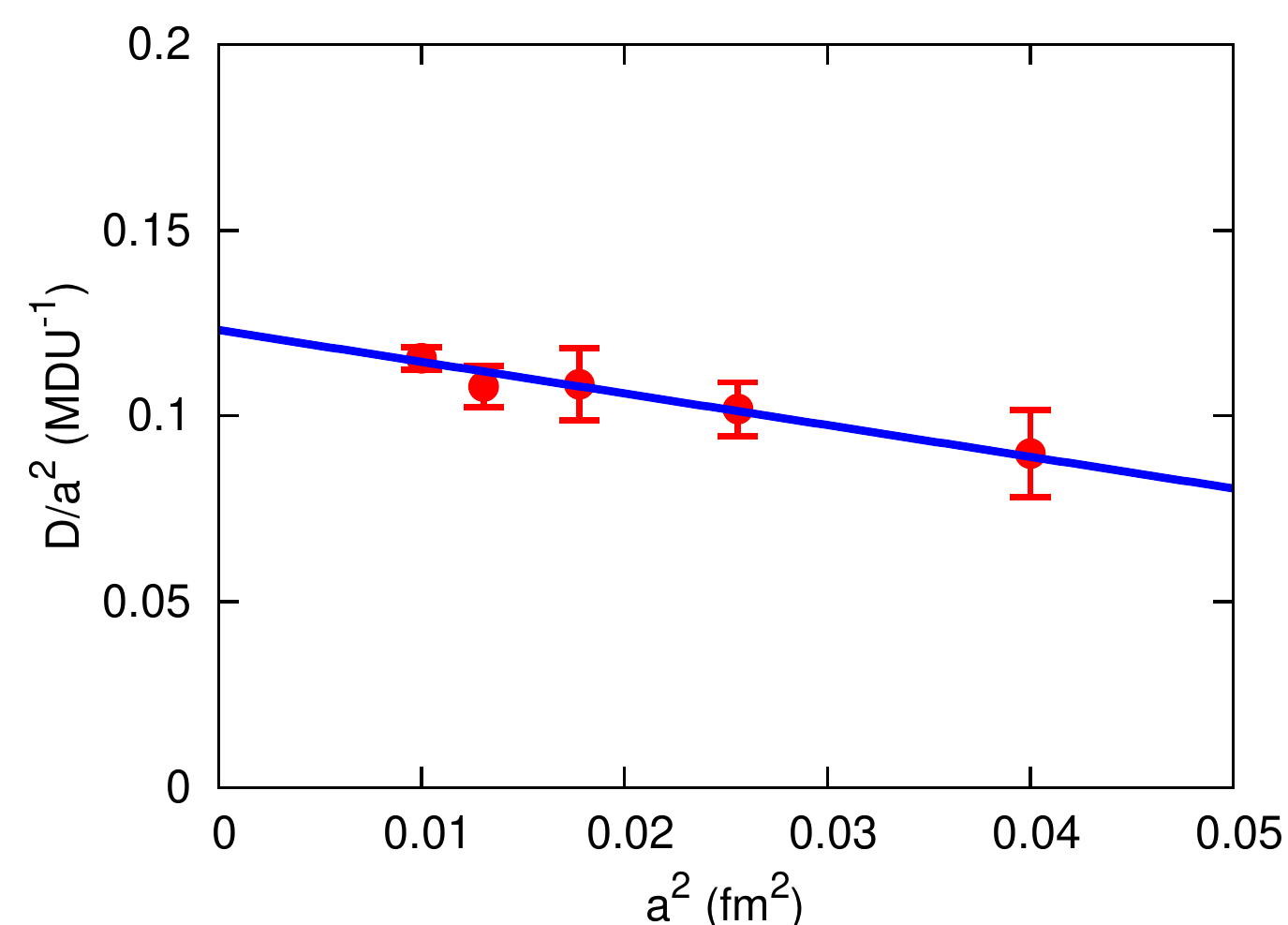}
\caption{Measured diffusion coefficient on periodic lattices versus $a^2$. The
linear fit is given by \refeq{eq:DFit}.}
\label{fig:D_vs_a}
\end{figure}

In Section \ref{sec:MeasAutocors} we gave some fits to the $a$-dependence of
$\tauint(Q)$, which is identical to the diffusion model parameter $\tautunn$.
It is very interesting to ask how the diffusion coefficient $D$, the other
parameter of the diffusion model, depends on the lattice spacing. The answer is
that over the range of lattice spacings we simulated $D$ scales like $a^2$ up
to small $O(a^4)$ corrections. Figure \ref{fig:D_vs_a} plots the fit results
for $D$ as a function of $a$ on periodic lattices, finding good agreement with
a fit of the form
\begin{equation} \label{eq:DFit} D/a^2 = c_1 + c_2 a^2, \,\,\,\,\, c_1 =
0.123(5) \text{ MDU}^{-1}, \,\, c_2 = -0.85(32) \text{ MDU}^{-1} \cdot \text{
fm}^{-2} \end{equation}

As we will discuss below in Section \ref{sec:TrajLenDep}, the parameters $D$
and $\tautunn$ depend on the parameters of the HMC algorithm, in particular the
trajectory length. In our simulations, we have chosen to scale the trajectory
length like $1/a$. It may be that a different choice for the scaling of the
trajectory length would lead to different scaling behaviors for $D$ and
$\tautunn$. In the rest of this paper we will assume that $D$ scales like
$a^2$, but it should be kept in mind that this could be modified to some extent
by difference choices for the scaling of the trajectory length.

The diffusion coefficient $D$ and the lattice Euclidean time extent $T$
together define a characteristic MD time we will call the ``diffusion
timescale,''
\begin{equation} \taudiff \equiv T^2/8D \end{equation}
With the factor of 8 included, this is roughly the MD time it takes a lump of
topological charge to diffuse across a distance $T/2$. It should be thought of
as the speed at which the center of the lattice can communicate with the
boundaries. From the scaling of $D$, the characteristic MD time interval
$\taudiff$ scales like $1/a^2$ at fixed Euclidean time extent $T$. 

\subsection{The tunneling- and diffusion-dominated regimes}

The diffusion model thus identifies a tunneling timescale $\tautunn$ and a
diffusion timescale $\taudiff$. There are two limiting cases where one of these
timescales is much shorter than the other.  In the ``tunneling-dominated''
regime characterized by $\tautunn \ll \taudiff$, diffusion is much slower than
tunneling, while in the ``diffusion-dominated'' regime where $\tautunn \gg
\taudiff$ diffusion is much faster than tunneling.

The tunneling-dominated regime corresponds to large $a$ (because then tunneling
is fast) or large $T$ (because then it takes a long time to diffuse across the
lattice). Conversely the diffusion-dominated regime corresponds to small $a$ or
small $T$. For a fixed physical value of $T$, coarse enough lattices will be
tunneling-dominated while fine enough lattices will be diffusion-dominated.
Similarly, for a fixed value of $a$, short enough lattices will be
diffusion-dominated while long enough lattices will be tunneling-dominated.

\begin{figure} \centering
\includegraphics[width=3.8in]{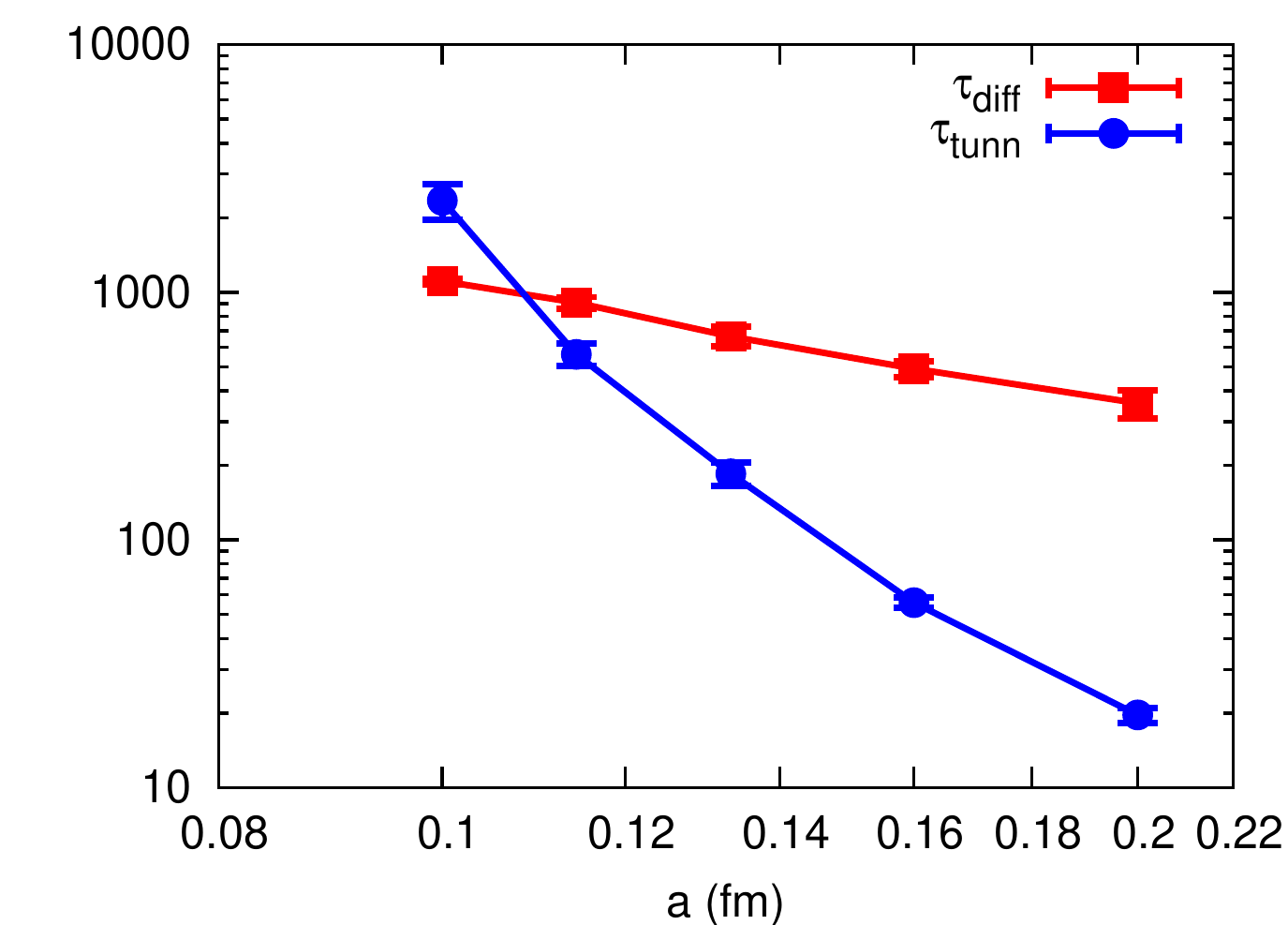}
\caption{Measurements of the tunneling and diffusion timescales from diffusion
model fits on all periodic lattices.}
\label{fig:Timescales}
\end{figure}

The transition region between these regimes is the region of parameter space
where $\tautunn \sim \taudiff$. Given the measurements of $\taudiff$ and
$\tautunn$ shown in Figure \ref{fig:Timescales}, this happens in our set of
ensembles at $a \sim 0.11$ fm. It should be kept in mind that the transition
between the tunneling- and diffusion-dominated regimes will happen at a
different lattice spacing depending on the action, Euclidean time extent, and
the HMC algorithm parameters. For example, if we used an action that was better
at topological tunneling, such as the Wilson gauge action, this transition
would occur at a finer lattice spacing.

\begin{figure} \centering
\includegraphics[width=5.3in]{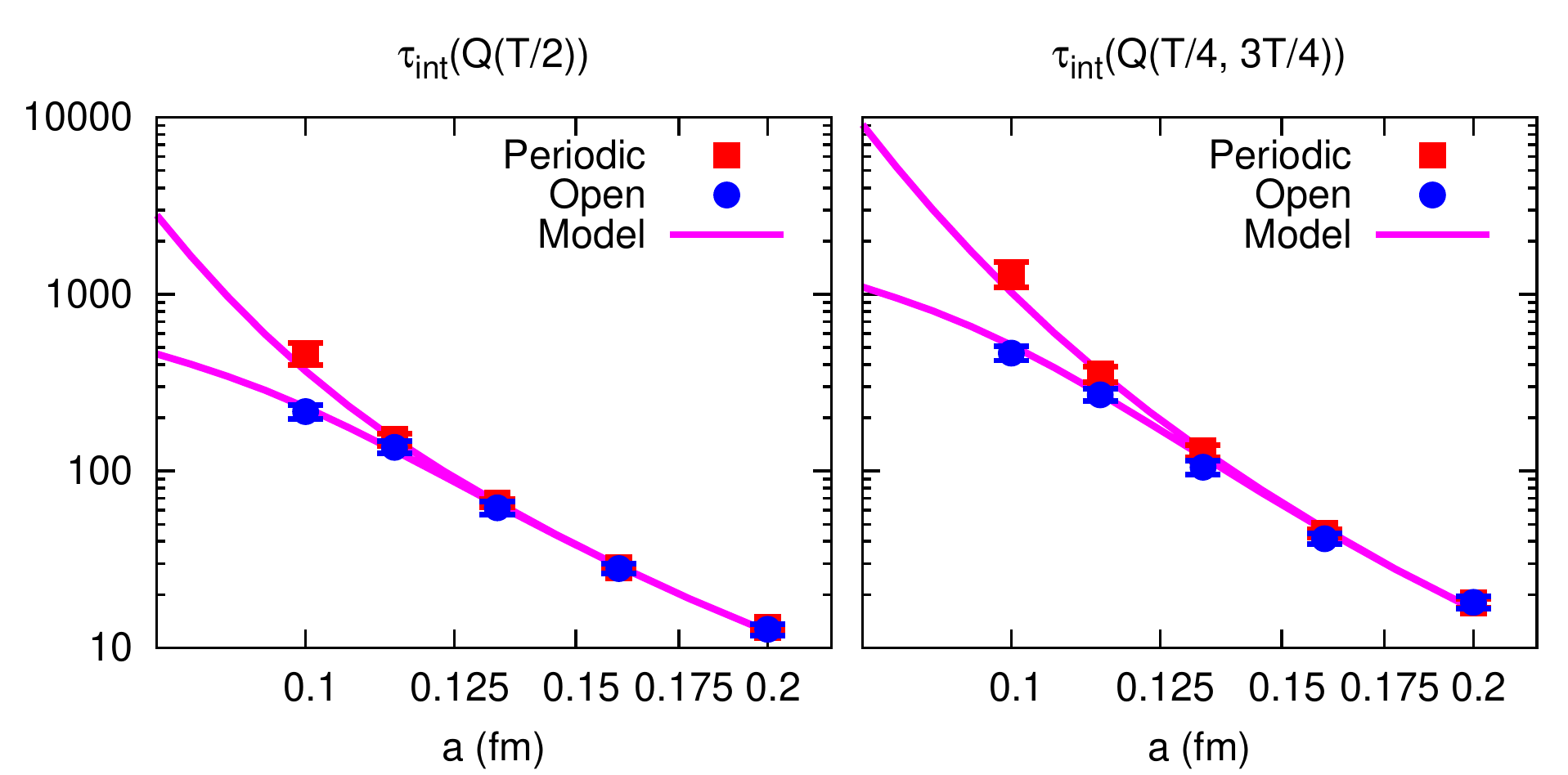}
\caption{Measurements and diffusion model predictions of $\tauint(Q(T/2))$
(time slice charge) and $\tauint(Q(T/4,3T/4))$ (half-volume charge) as a
function of the lattice spacing.}
\label{fig:SubvolumeIATs_vs_a}
\end{figure}

Tunneling happens at equal rates on periodic and open lattices. Therefore in
the tunneling-dominated regime there will be little difference between the
autocorrelation times on periodic and open lattices. Because $\tautunn$ is so
short, autocorrelations are destroyed by tunneling much faster than the
timescale $\taudiff$ on which the boundaries can affect the bulk. In the
diffusion-dominated regime, however, we expect significant differences between
autocorrelation times on open and periodic lattices. On open lattices, in the
diffusion-dominated regime, the topological charge in the bulk can change even
in the absence of tunneling by exchanging topological charge with the
boundaries, where topological charge can be created and destroyed freely.

Figure \ref{fig:SubvolumeIATs_vs_a} shows that the integrated autocorrelation
times of the time slice charge $Q(T/2)$ and the half-volume charge
$Q(T/4,3T/4)$ both follow this pattern. At coarse lattice spacings, periodic
and open boundary conditions produce identical autocorrelation times. At fine
lattice spacings, open boundary conditions produce much shorter autocorrelation
times. The transition region between these two regimes indeed occurs at $a
\approx 0.11$ fm. 

Also plotted in Figure \ref{fig:SubvolumeIATs_vs_a} are the integrated
autocorrelation times for these observables calculated with the diffusion
model, using as inputs the measured scaling behavior of $\tautunn$ and $D$ (we
have neglected the boundary region enhancement of $D(t)$ on open lattices,
which has only a minor effect on these integrated autocorrelation times). The
model curves correctly reproduce the observed behavior and show that if we ran
simulations at even finer lattice spacings the difference between periodic and
open boundary conditions would become very large. 

How exactly do autocorrelation times scale with $a$? How much do open boundary
conditions improve the scaling the diffusion-dominated regime? In principle to
answer these questions all we have to do is numerically integrate
\refeq{eq:DiffEq}, making use of our knowledge of the simple $a$-dependence of
$\tautunn$ and $D$. That is what we have done in Figure
\ref{fig:SubvolumeIATs_vs_a}. However, we can do better: in the tunneling- or
diffusion-dominated limits we can obtain analytic results for some integrated
autocorrelation times in the diffusion model. We work these out in the next
section.

\subsection{Analytic scaling laws in the tunneling- and diffusion-dominated
regimes} \label{sec:AnalyticScaling}

In this section we use \refeq{eq:DiffEq} to compute the integrated
autocorrelation time of $Q(t_0)$, the topological charge on the time slice at
Euclidean time $t_0$, in the tunneling- and diffusion-dominated regimes on
periodic and open lattices. We will obtain analytic results telling us how this
integrated autocorrelation time scales with $a$ in each of these limits. While
we will carry out the computations for the specific observable $Q(t_0)$, some
of the results will generalize to all topological observables.

In these calculations we make a few simplifications to make the problem more
analytically tractable. First, we will treat Euclidean time as continuous, so
that the $\partial/\partial t$'s in \refeq{eq:DiffEq} will be actual
derivatives instead of finite differences. This is a good approximation because
$C(t, t_0, \tau)$ is always fairly smooth as a consequence of the Wilson flow
smearing that goes into measuring $Q(t)$. Second, we will ignore the fact that
in real simulations measurements are only conducted at discrete MD times
separated by an interval $\taumeas$; we will simply compute the integrated
autocorrelation time as an integral:
\begin{equation} \label{eq:TauIntIntegral} \tauint(Q(t_0)) \equiv
\int_0^{\infty} d\tau \rho_{Q(t_0)}(\tau) = \int_0^{\infty} d\tau \frac{C(t_0,
t_0, \tau)}{C(t_0, t_0, 0)} \end{equation}
whereas in an actual simulation $\tauint$ would have to be computed as a
discrete sum. Finally, we will ignore the $t$-dependence of $D$ near the open
boundaries, which will not change the qualitative conclusions as long as $T$ is
significantly larger than the boundary region in which $D(t)$ is not constant.

In order to make predictions we need as input the equal-MD-time correlation
function $C(t, t_0, 0)$. In our simulations, we find that $C(t, t_0, 0)$ is
very close to Gaussian:
\begin{equation} \label{eq:GaussianInitCond} C(t, t_0, 0) \approx c
e^{-(t-t_0)^2/2\sigma^2} \end{equation}
where $\sigma$ is a physical length scale which we find to be about 0.22 fm.
The scaling predictions in this section use this Gaussian form for $C(t, t_0,
0)$. However the exact shape of $C(t, t_0, 0)$ is not important for the
qualitative conclusions we will draw about scaling. 

With these simplifications the problem amounts to solving the simple linear
differential equation \refeq{eq:DiffEq} with initial conditions given by
\refeq{eq:GaussianInitCond} and then performing the integral in
\refeq{eq:TauIntIntegral}. We relegate the details to an appendix and just give
the results here.

\subsubsection{The tunneling-dominated regime} 

In the tunneling dominated regime, autocorrelation times are independent of the
boundary conditions for observables located far enough from the boundaries.
Here ``far enough'' means a distance greater than about $\sqrt{2 D \tautunn}$.
If $t_0$ satisfies this condition, we find
\begin{equation} \label{eq:TauIntTunnLimit} \tauint(Q(t_0)) \approx
\sqrt{\frac{\pi \sigma^2 \tautunn}{2 D}} \end{equation}
where the approximation is good up to corrections down by powers of
$\sigma/\sqrt{2D\tautunn}$. These corrections become small at fine enough
lattice spacings, well before the lattice spacing becomes fine enough that we
transition from the tunneling-dominated to the diffusion-dominated regime.
\refeq{eq:TauIntTunnLimit} says that in the tunneling-dominated regime this
integrated autocorrelation time scales essentially like $\sqrt{\tautunn / D}$.
This scaling is not quite as bad as that of $\tautunn$ itself but it is still
quite bad. We will now see that the scaling in the diffusion-dominated regime
is worse than this on periodic lattices, but better than this on open lattices.

\subsubsection{The diffusion-dominated regime on periodic lattices}

On a periodic lattice, the large-$\tau$ behavior of the normalized
autocorrelation function of $Q(t_0)$ is:
\begin{equation} \rho_{Q(t_0)}(\tau) \xrightarrow{\tau \to \infty}
\sqrt{2\pi}\frac{\sigma}{T} e^{-\tau/\tautunn} \end{equation}
In the diffusion-dominated limit, $\tautunn$ is very large and $\tauint$
becomes dominated by the area under this tail, so that
\begin{equation} \label{eq:TauIntDiffPeriodic} \tauint(Q(t_0)) \to \sqrt{2 \pi}
\frac{\sigma}{T} \tautunn \end{equation}
So on periodic lattices in the diffusion-dominated regime, $\tauint(Q(t_0))$
scales in the same (very bad) way as $\tautunn$, the integrated autocorrelation
time of the global topological charge.

This result generalizes beyond the time-slice charge: in the
diffusion-dominated limit on a periodic lattice all topological autocorrelation
times scale like $\tautunn$, and thus increase very rapidly as $a \to 0$. The
reason is that all topological autocorrelation functions develop long tails of
the form $\exp(-\tau/\tauexp)$ with $\tauexp \propto \tautunn$, and for
$\tautunn$ large enough the area under this tail dominates the integrated
autocorrelation time.

\subsubsection{The diffusion-dominated regime on open lattices}

On an open lattice in the diffusion-dominated limit, we find
\begin{equation} \label{eq:TauIntDiffOpen} \tauint(Q(t_0)) \approx  \sqrt{2
\pi} K \frac{t_0}{T} \left(1 - \frac{t_0}{T}\right) \frac{\sigma T}{D}
\end{equation}
where $K = 1 + O(\sigma/T)$ is a near-unity coefficient. This formula gives the
form of the $t_0$-dependence of $\tauint(Q(t_0))$ on open lattices, although it
should be kept in mind that the exact form of the $t_0$-dependence will be
modified by the time-dependence of the diffusion coefficient near the
boundaries, which we have neglected in this calculation. 

\refeq{eq:TauIntDiffOpen} shows that in the diffusion-dominated regime
$\tauint(Q(t_0))$ scales in the same way as $1/D$.  Above we found that $D$
scales like $a^2$, so in this regime the integrated autocorrelation time of
$Q(t_0)$ scales like $1/a^2$. In fact this scaling behavior generalizes to all
topological autocorrelation times. This is because in the diffusion-dominated
limit on open lattices the term in \refeq{eq:DiffEq} proportional to
$1/\tautunn$ can be dropped. Then any quantity with units of MD time that we
can construct from the parameters of the diffusion model is proportional to
$1/D$. Thus in the diffusion-dominated limit on open lattices, all topological
autocorrelation times scale like $1/a^2$.

Finally, we see that at fixed $t_0/T$ this integrated autocorrelation time is
proportional to $T$. So while the $1/a^2$ scaling is an improvement over the
scaling seen with periodic boundary conditions, if the lattice has a large
Euclidean time extent the coefficient in front of the $1/a^2$ will be large.

\subsection{More general topological observables}

\begin{figure} \centering
\includegraphics[width=3.8in]{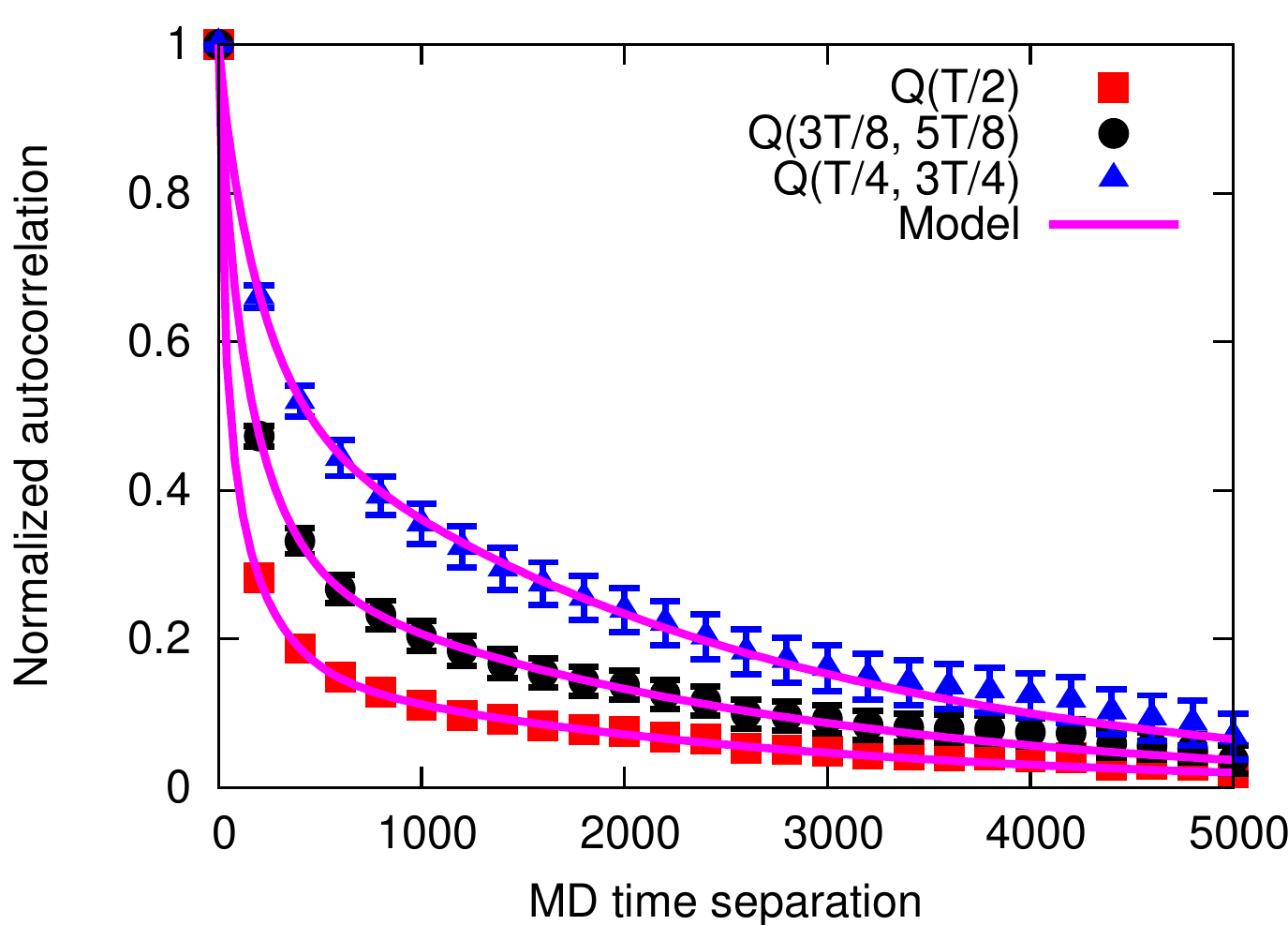}
\caption{Measurements and model predictions of the normalized autocorrelation
functions of the topological charge summed over the central time slice, central
quarter-volume, and central half-volume on the periodic $a = 0.100$ fm
ensemble.}
\label{fig:SubvolumeACFs}
\end{figure}

The mathematical model we have given describes the MD-time correlation
functions of the time-slice topological charges $Q(t)$. We can combine these to
find the autocorrelation function of $Q(t_1, t_2)$, the charge summed over a
finite Euclidean time extent $t_1 \le t < t_2$ by expressing the subvolume
charge as a sum of time slice charges:
\begin{equation} \label{eq:SubvolumeACF} \langle Q(t_1, t_2, \tau_1) Q(t_1,
t_2, \tau_2)\rangle = \sum_{t_a = t_1}^{t_2-a} \sum_{t_b = t_1}^{t_2-a} \langle
Q(t_a, \tau_1) Q(t_b, \tau_2) \rangle \end{equation}
where here $Q(t_1, t_2, \tau)$ is the value of the subvolume charge $Q(t_1,
t_2)$ at the MD time $\tau$. The diffusion model gives all the correlation
functions on the right-hand side, so we can use it to compute the left-hand
side also. For example, in Figure \ref{fig:SubvolumeACFs} we plot the
normalized autocorrelation functions of several subvolume charges on the
periodic $a = 0.100$ fm ensemble and demonstrate that they are in close
agreement with the model predictions calculated using \refeq{eq:SubvolumeACF}.

We can also use the model to compute the autocorrelations of squared charges
like $Q(t)^2$ and $Q(t_1, t_2)^2$. As noted in \cite{LS_OBC}, if $X$ is any
observable defined by a sum over a region of the lattice much larger than the
longest physical QCD correlation length, then there is a simple relationship
\begin{equation} \rho_{X^2}(\tau) = \rho_X(\tau)^2 \end{equation}
between the normalized autocorrelation function of $X$ and the normalized
autocorrelation function of $X^2$. Thus for instance we can predict the
normalized autocorrelation function of $Q(t)^2$ just by squaring the prediction
for the normalized autocorrelation function of $Q(t)$.

\subsection{Dependence of diffusion model parameters on HMC algorithm
parameters}
\label{sec:TrajLenDep}

Autocorrelation times are properties of the HMC algorithm and not properties of
the simulated theory alone. Therefore the tunneling and diffusion timescales
that we have defined may depend not just on the lattice spacing and the
Euclidean time extent but also on the parameters of the simulation algorithm.
Here the only parameter we will consider is the trajectory length.

\begin{figure} \centering
\includegraphics[width=5.3in]{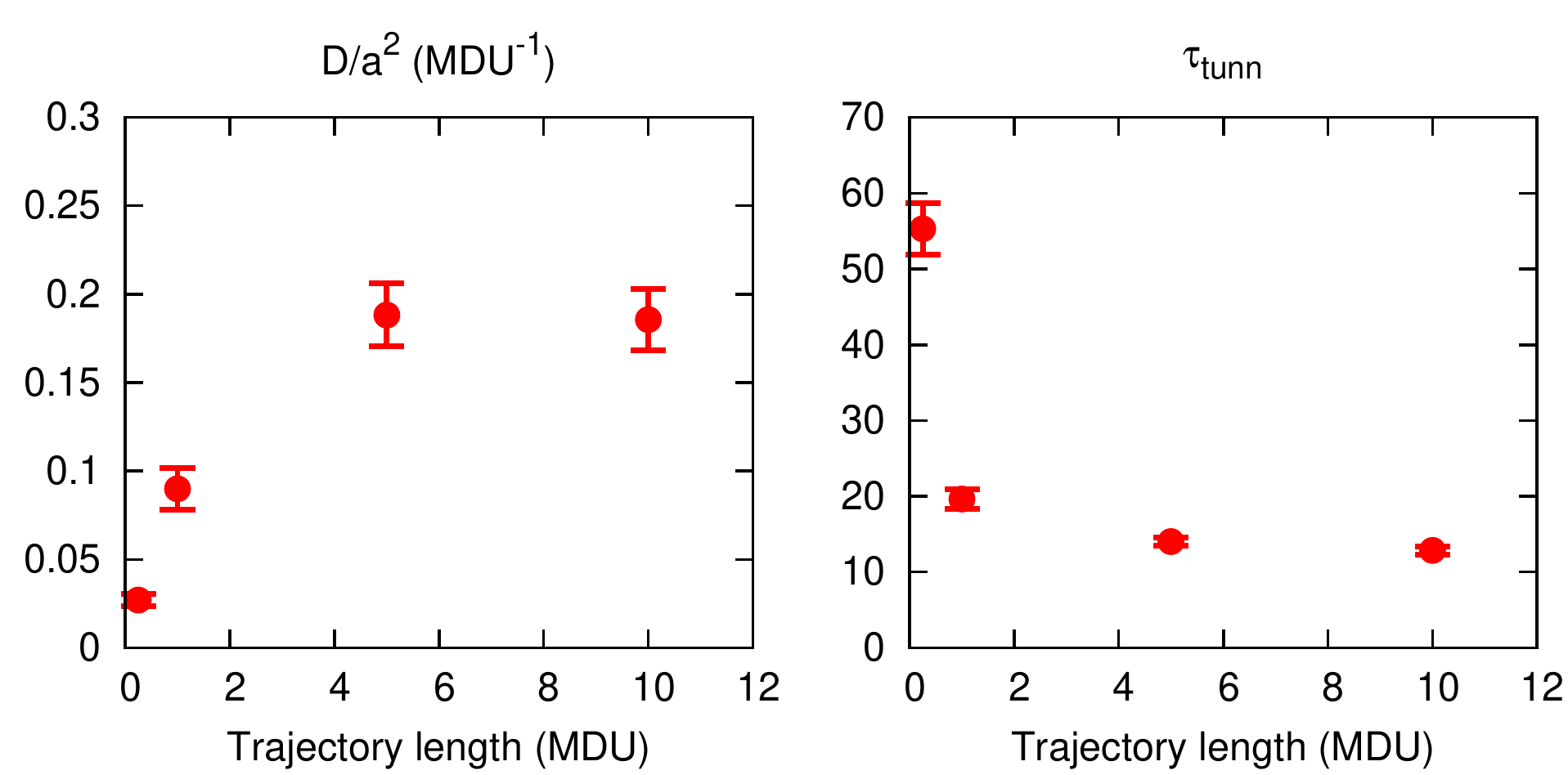}
\caption{Fits to the diffusion model parameters $D$ and $\tautunn$ on four
versions of the periodic $a = 0.200$ fm ensemble with four different trajectory
lengths $\tau_{\rm traj} = 0.25, 1, 5, 10$ MDU. }
\label{fig:TrajlenDependence}
\end{figure}

We ran several additional simulations with different HMC trajectory lengths at
our coarsest lattice spacing in order to measure the influence of the
trajectory length on the diffusion coefficient $D$ and the tunneling timescale
$\tautunn$. The measurement interval $\taumeas = 10$ MDU was held constant as
the trajectory length was varied, and the MD integrator step size was adjusted
to keep acceptance high. As shown in Figure \ref{fig:TrajlenDependence}, we
find that longer trajectories lead to larger $D$ and smaller $\tautunn$. This
is consistent with previous experience which suggests that increasing the
trajectory length can decrease autocorrelation times \cite{AlphaTrajLen,
Alpha}.

\section{Conclusions} \label{sec:Conclusions}

We have shown that the autocorrelation functions of topological observables are
predicted very accurately by a surprisingly simply mathematical model that
incorporates only two processes: diffusion of topological charge and tunneling
between topological sectors. The rates of these processes are given by the
diffusion coefficient $D$ and the tunneling timescale $\tautunn$ (which on a
periodic lattice is just the integrated autocorrelation time of the global
topological charge). We find that $D$ scales like $a^2$ while $\tautunn$ scales
like $\exp(k \beta)$ with $k$ uncomfortably large.

The relative rates of the tunneling process and the diffusion process determine
whether open boundary conditions are useful for reducing autocorrelation times.
The characteristic timescale of the diffusion process is $\taudiff = T^2/8D$.
Open boundary conditions show drastically improved scaling of autocorrelation
times in the diffusion-dominated regime, when
\begin{equation} \label{eq:DiffDom} T^2/8D \ll \tautunn \end{equation}
In this regime, topological autocorrelation times scale like $1/D \sim 1/a^2$
on open lattices, while on periodic lattices they are proportional to
$\tautunn$. It is when \refeq{eq:DiffDom} is satisfied that open boundary
conditions are useful for reducing autocorrelation times.

In the opposite limit of $T^2/8D \gg \tautunn$ the simulation is
tunneling-dominated. In this regime autocorrelation times in the bulk are
independent of the boundary conditions and open boundary conditions will not
reduce autocorrelation times.

As an example of applying this criterion, we can consider the Wilson gauge
action simulations of \cite{MMProc}, mentioned in the introduction, in which we
attempted to compare open and periodic boundary conditions with the Wilson
gauge action at $\beta = 6.42$, $a \sim 0.05$ fm, $T/a = 32$. The statistics of
those simulations were quite low given the long autocorrelations, but we can
estimate the order of magnitude of the tunneling timescale $\tautunn$ from the
high-statistics simulations of \cite{Alpha}. There a simulation with the same
gauge action (but a slightly different simulation algorithm) at the nearby
value of $\beta = 6.475$ found the integrated autocorrelation time of $Q^2$ to
be several thousand MDU; this gives the right order of magnitude for
$\tautunn$, which is the integrated autocorrelation time of $Q$.  Meanwhile if
the diffusion constant for the Wilson gauge action is similar to the value
$D/a^2 \sim 0.1 \text{ MDU}^{-1}$ measured here with the DBW2 action, then the
diffusion timescale is $\taudiff = T^2/8D \sim 1000$ MDU. The situation is
therefore likely similar to the DBW2 lattices at $a = 0.100$ fm in this paper:
$\taudiff$ is smaller than $\tautunn$ by a factor of a few, so open boundary
conditions should decrease autocorrelation times by a factor of a few. However
both timescales are quite long, and with the limited statistics we collected in
\cite{MMProc} we were not able to measure the long autocorrelations accurately
enough to detect this difference.

\refeq{eq:DiffDom} will be satisfied eventually for small enough $a$, so open
boundary conditions will always be useful if the lattice spacing is fine
enough. How small $a$ needs to be depends on the lattice action (which controls
the tunneling timescale) and the Euclidean time extent. The faster the
tunneling timescale and the longer the lattice, the finer the lattice spacing
needs to be before open boundary conditions will be useful. Long lattices have
an additional drawback: in the diffusion-dominated regime on open lattices some
integrated autocorrelation times are proportional to $T$.

\section{Acknowledgements}

We thank Martin L\"uscher and Stefan Schaefer as well as our fellow members of
the RBC collaboration for useful conversations. We thank RIKEN-BNL Research
Center and Brookhaven National Lab for the use of the Blue Gene/Q computers.
This work was supported by US DOE grant \#DE-FG02-92ER40699. 

\begin{appendix}
\section{Appendix: Analytic calculation of $\tauint(Q(t_0))$}
\label{sec:AnalyticAppendix}

Here we supply some of the details in the calculations of $\tauint(Q(t_0))$
in Section \ref{sec:AnalyticScaling}.

For convenience we will scale $C(t, t_0, 0)$ so that $C(t_0, t_0, 0) = c = 1$;
then the normalized autocorrelation function of $Q(t_0)$ is
$\rho_{Q(t_0)}(\tau) = C(t_0, t_0, \tau)$.

\subsection{The tunneling-dominated regime} 

In the tunneling-dominated regime, as long as $t_0$ is not too close to an open
boundary we can pretend that we are working with a lattice of infinite
Euclidean time extent, because the correlations measured by $C(t, t_0, \tau)$
are destroyed by tunneling before they can diffuse to the boundaries of the
lattice. Solving \refeq{eq:DiffEq} on an infinite domain with initial
conditions given by \refeq{eq:GaussianInitCond} we obtain
\begin{equation} C(t, t_0, \tau) = \frac{1}{\sqrt{1 + 2 D \tau / \sigma^2}}
\exp \left( \frac{-(t-t_0)^2}{2\sigma^2 + 4D\tau} -
\frac{\tau}{\tautunn}\right) \end{equation}
The normalized autocorrelation function of $Q(t_0)$ is then
\begin{equation} \rho_{Q(t_0)}(\tau) = C(t_0, t_0, \tau) =
\frac{e^{-\tau/\tautunn}}{\sqrt{1 + 2D\tau/\sigma^2}} \end{equation}
and the integrated autocorrelation time is
\begin{equation} \tauint(Q(t_0)) = \int_0^\infty d\tau \rho_{Q(t_0)}(\tau) =
\tautunn \left[ \sqrt\pi x - 2 x^2 + \sqrt\pi x^3 + O(x^4) \right]
\end{equation}
where $x = \sigma/\sqrt{2D\tautunn}$.

\subsection{The diffusion-dominated regime on periodic lattices}

At finite $T$, \refeq{eq:DiffEq} can be solved by solving the eigenvalue
equation 
\begin{equation} \label{eq:EigenvalueEq} \left(-D \frac{\partial^2}{\partial
t^2} + \frac{1}{\tautunn}\right)\phi_n(t) = \lambda_n \phi_n(t) \end{equation}
where the $\phi_n$ satisfy periodic boundary conditions, and then expanding
$C(t, t_0, \tau)$ in the eigenmodes $\phi_n$ as
\begin{equation} C(t, t_0, \tau) = \sum_n c_n \phi_n(t) e^{-\lambda_n \tau}
\end{equation}
The integrated autocorrelation time of $Q(t_0)$ is then 
\begin{equation} \label{eq:TauIntEigenmodes} \tauint(Q(t_0)) = \sum_n
\frac{c_n}{\lambda_n} \phi_n(t_0) \end{equation}

Without loss of generality we will take $t_0 = 0$. The eigenfunctions and
eigenvalues are 
\begin{equation} \phi_n(t) = \cos\left(\frac{2n\pi t}{T}\right), \,\,\,\,
\lambda_n = \frac{1}{\tautunn} + D \left(\frac{2 n \pi }{T}\right)^2 , \,\,\,\,
n = 0, 1, 2, ...\end{equation}
where we have ignored the odd eigenfunctions of \refeq{eq:EigenvalueEq}
because they are orthogonal to the initial state $C(t, 0, 0)$. In the
diffusion-dominated limit, $\tautunn$ becomes large and so $\lambda_0 =
1/\tautunn$ becomes much smaller than all the other eigenvalues. In this limit,
the normalized autocorrelation function develops a long tail at large $\tau$:
\begin{equation} \rho_{Q(t_0)}(\tau) = C(t_0, t_0, \tau) \xrightarrow{\tau \to
\infty} \sqrt{2\pi}\frac{\sigma}{T} e^{-\tau/\tautunn} \end{equation}

In the diffusion-dominated limit, $\tauint$ becomes dominated by the area under
this tail, so that
\begin{equation} \tauint(Q(t_0)) = \frac{c_0}{\lambda_0} +
O\left(\frac{1}{\lambda_1}\right) = \sqrt{2 \pi} \frac{\sigma}{T}\tautunn +
O\left(\frac{T^2}{4\pi^2 D}\right) \end{equation}

\subsection{The diffusion-dominated regime on open lattices}

On open lattices we again must solve the eigenvalue problem
\refeq{eq:EigenvalueEq} but this time the boundary conditions are $\phi_n(0) =
\phi_n(T) = 0$. The eigenfunctions and eigenvalues are 
\begin{equation} \label{eq:EigenvaluesOpen} \phi_n(t) = \sin\left(\frac{n \pi
t}{T}\right), \,\,\,\, \lambda_n = \frac{1}{\tautunn} + D \left( \frac{n
\pi}{T} \right)^2 , \,\,\,\, n = 1, 2, ...\end{equation}
This time there is no near-zero eigenvalue when $\tautunn$ becomes large, so
the sum in \refeq{eq:TauIntEigenmodes} is not dominated by a single mode.
Therefore in the diffusion-dominated limit we can drop the $1/\tautunn$ term in
\refeq{eq:EigenvaluesOpen}. For $n \ll T/\sigma$, a good approximation is
\begin{equation} c_n \approx \sqrt{8 \pi} \frac{\sigma}{T} \sin\left(\frac{n
\pi t_0}{T}\right) \end{equation}
For $n > T/\sigma$, $c_n$ goes rapidly to zero. Then using
\refeq{eq:TauIntEigenmodes} we can write a good approximation to $\tauint$:
\begin{equation} \tauint(Q(t_0)) \approx  \sqrt{8 \pi}\frac{\sigma T}{\pi^2 D}
\sum_{n=1}^{T/\sigma} \frac{1}{n^2} \sin^2 \left(\frac{n \pi t_0}{T}\right)
\end{equation}
We can extend this finite sum to an infinite sum at the cost of an
$O(\sigma/T)$ error. Then using the fact that, for $a \in [0, \pi]$,
\begin{equation} \sum_{n=1}^\infty \frac{\sin^2 (a n)}{n^2} =
\frac{1}{2}a(\pi-a) \end{equation}
we get 
\begin{equation} \tauint(Q(t_0)) \approx \sqrt{2 \pi} K \frac{t_0}{T} \left(1 -
\frac{t_0}{T}\right) \frac{\sigma T}{D} \end{equation}
where $K$ is some coefficient of order $1 + O(\sigma/T)$ that accounts for the
error introduced by going from a finite sum to an infinite one.

\end{appendix}

\end{document}